\documentclass{ar-1col-S2O}[a4paper]

\usepackage[comma]{natbib}
\usepackage{url}
\usepackage{xcolor}
\usepackage{amssymb}
\usepackage{amsmath}
\usepackage{subfig}

\jname{Annu.\ Rev.\ Astron.\  Astrophys.}
\jvol{AA}
\jyear{2022}

\begin{document}
\newcommand{\HI}{H\,{\sc i}}
\newcommand{\htwo}{H$_2$}
\newcommand{\halpha}{H${\rm \alpha}$}
\newcommand{\MHI}{${\rm M_{\rm HI}}$}
\newcommand{\Msun}{$\rm {M_{\odot}}$}

\newcommand\arcdeg{\mbox{$^\circ$}}%
\newcommand\arcmin{\mbox{$^\prime$}}%
\newcommand\arcsec{\mbox{$^{\prime\prime}$}}%
\newcommand\fd{\mbox{$.\!\!^{\mathrm d}$}}%
\newcommand\fh{\mbox{$.\!\!^{\mathrm h}$}}%
\newcommand\fm{\mbox{$.\!\!^{\mathrm m}$}}%
\newcommand\fs{\mbox{$.\!\!^{\mathrm s}$}}%
\newcommand\fdg{\mbox{$.\!\!^\circ$}}%

\newcommand{\bighicat}{{\sc bighicat}}

\newcommand{\kms}{km\,s$^{-1}$}
\newcommand{\aaps}{Astron.\ \& Astrophys.\ Suppl.\ Ser.}%
\newcommand{\apss}{Astrophys.\ \& Space Sci.}%

\newcommand{\aj}{Astron.\ J.}%
\newcommand{\apj}{Astrophys.\ J.}
\newcommand{\apjs}{Astrophys.\ J.\ Suppl.}
\newcommand{\apjl}{Astrophys.\ J.\ Lett.}
\newcommand{\actaa}{Acta Astronom.}
\newcommand{\aap}{Astron.\ \& Astrophys.}
\newcommand{\araa}{Ann.\ Rev.\ Astron. Astrophys.}
\newcommand{\mnras}{Mon.\ Not.\ R.\ Astron.\ Soc.}%
\newcommand{\nar}{New Astron.\ Rev.}%
\newcommand{\pasa}{Publ.\ Astron.\ Soc.\ Aust.}%
\newcommand{\pasp}{Publ.\ Astron.\ Soc.\ Pacific}%
\newcommand{\pasj}{Publ.\ Astron.\ Soc.\ Japan}%
\newcommand{\nat}{Nature}%

\newcommand{\rh}[1]{\textcolor{red}{#1}}

\newcommand{\naomi}[1]{\textcolor{orange}{Naomi:~ #1}}
\newcommand{\snez}[1]{\textcolor{blue}{Snez: #1}}
\newcommand{\dan}[1]{\textcolor{violet}{Dan: #1}}

\markboth{McClure-Griffiths, Stanimirovi\'c, \& Rybarczyk}{\HI\ in the Milky Way}

\title{Atomic Hydrogen in the Milky Way: A Stepping Stone in the  Evolution of Galaxies }
\author{Naomi M. McClure-Griffiths,$^1$ Sne\v{z}ana Stanimirovi\'c$^2$ and Daniel R. Rybarczyk$^2$
\affil{$^1$Research School of Astronomy \& Astrophysics, Australian National University, Canberra ACT, Australia 2610  email: naomi.mcclure-griffiths@anu.edu.au}
\affil{$^2$Department of Astronomy, University of Wisconsin--Madison, Madison WI, USA}
}
\begin{abstract}Atomic hydrogen (\HI) is a critical stepping stone in the gas evolution cycle of the interstellar medium (ISM) of the Milky Way. \HI\ traces both the cold, pre-molecular state before star-formation and the warm, diffuse ISM before and after star-formation.  This review describes new, sensitive \HI\ absorption surveys, which together with high angular and spectral resolution \HI\ emission data,  have revealed the physical properties of \HI, its structure and its association with magnetic fields.  We give an overview of the \HI\ phases, discuss how \HI\ properties depend on environment and what its structure can tell us about feedback in the ISM.  Key findings include:

\vspace{0.2cm}
\begin{minipage}{10cm}
\begin{itemize}
\item The mass fraction of the cold neutral medium is $\lesssim 40$\% on average, increasing with $A_V$ due to the increase of mean gas density.
\item The cold disk extends to at least $R\sim 25$ kpc.
\item Approximately 40\% of the \HI\ is warm with structural characteristics that derive from feedback events.
\item Cold \HI\ is highly filamentary, whereas warm \HI\ is more smoothly distributed.
\end{itemize}
\end{minipage}

\vspace{0.2cm}
 We summarize future observational and simulation opportunities that can be used to unravel the 3-D structure of the atomic ISM and the effects of heating and cooling on \HI\ properties. 
\end{abstract}

\begin{keywords}
atomic hydrogen (\HI), interstellar medium, galaxy evolution, magnetic fields 

\end{keywords}
\maketitle

\tableofcontents

\section{INTRODUCTION}
	
Neutral hydrogen (\HI) gas is the fundamental building  block of galaxies.   In the local Universe fully three-quarters of the neutral gas of galaxies is in the atomic phase \citep{carilli13}.    \HI\  defines a galaxy's morphology on the evolutionary path between the ionized intergalactic medium and the molecular clouds where stars are formed.   
\HI\ is observed to have structure on all spatial scales, from tiny, AU scales observed in the Milky Way \citep[reviewed in][]{stanimirovic18}, to Galactic-scale coherent structures \citep[e.g.][]{mcgriff04}.  And, unlike many standard tracers of the structure of a galaxy, \HI\ is not biased towards tracing star formation - it follows spiral arms with and without stars, probes gaseous disks extending beyond the stellar disk \citep{saintonge22} and even traces the interaction between a galaxy and its starless circumgalactic environment \citep{cortese21}.  

Around galaxies, \HI\ exists in the form of highly diffuse, infalling, gas that replenishes the disk \HI\ reservoir and provides an important energy contribution. Cosmological simulations predict diffuse, infalling \HI\ with column density $<10^{18}$ cm$^{-2}$, and large concentrations originating from galaxy interactions \citep{stevens19}.  Around the Milky Way, large concentrations of \HI\ are  associated with the Magellanic System  and smaller, dense concentrations are scattered throughout the halo and disk-halo interface as High and Intermediate Velocity Clouds.  In this review, we will not focus on the Milky Way's circumgalactic \HI, which is well reviewed by \cite{donghia16} and  \cite{putman12}.

Instead, we will focus on the properties of \HI\ in the disk of the Milky Way where it is perpetually in the act of ``flowing" between its dense, cold state and its warm, diffuse state.  \HI\ is present over a range of temperatures between $\sim 20~{\rm K}$ and $\sim 8000~{\rm K}$ \citep{heiles03b}, tracing most of the semi-quiescent conditions of a galaxy.    While the star formation rate of a galaxy is tightly linked to its molecular hydrogen (H$_2$) content,  the star formation rate per unit \HI\ mass varies widely among and within galaxies \citep{leroy08}. The standard interpretation is that the star formation in galaxies is separated into two processes:  one in which molecular clouds are assembled from \HI\ and the second in which stars form from H$_2$ \citep{schruba11}.  In fact, the key regulator of star formation appears not to be the formation of H$_2$ itself, but rather the formation of gas that is sufficiently shielded to reach very low temperatures so that thermal pressure cannot prevent collapse \citep{krumholz11,glover12}. \HI\ can therefore be considered the fuel for star formation and the key ingredient responsible for the shielding and survival of molecular gas.  The transition from warm, unshielded to cold, UV-shielded \HI\ might act as a  throttle in the process of converting the interstellar medium (ISM) into stars, making it essential to the evolution of galaxies.  


In this review we will expand on the role of \HI\ in a galaxy  using the Milky Way as the template.  We particularly focus on the properties of the cold neutral medium (CNM) to highlight the first step in the gas evolution cycle from diffuse, warm to cold and molecular.   Over the last 40 years, \HI\ has been primarily studied in emission and the focus has been on the large-scale structure and morphology of the warm neutral medium (WNM).  Recent improvements in observational techniques for \HI\ absorption, together with improved spatial and spectral resolution revealing cold, small-scale filamentary and fibrous structures in emission, have redirected Galactic \HI\ studies towards the CNM as a stepping stone towards molecular cloud and star formation.   Whereas the WNM is smooth and ubiquitous, the CNM has abundant small-scale structure and is better observed on small scales that are easily accessed in the Milky Way.  Only in the Milky Way can we observe \HI\ with sufficient resolution to probe the physical scales that are relevant to the cooling, condensing, and turbulent dissipation that affect how \HI\ converts to H$_2$, as well the feedback scales on which \HI\ is heated and turbulence driven.   
Understanding the CNM as a phase of \HI\ is essential for understanding \HI\ as a path to molecular clouds and star formation.


The topic of \HI\ in the Milky Way disk is vast and has been periodically reviewed \citep[e.g.,][]{kulkarni87,burton88,dickey90,kalberla09}.  We strive to use this review to emphasize the importance of \HI\ in understanding how galaxies evolve
and what \HI\ can and cannot tell us about the ISM.  We will describe what has been revealed about the fundamental properties of \HI\ over the last decade, what we still don't know, and finally the prospects for the field over the coming decade.  While most of us learn about the hyper-fine transition of atomic hydrogen at 1420.405752 MHz in our undergraduate physics classes, few learn about how it is measured and the radiative transfer that leads to its widespread usage as a tracer of mass and dynamics within galaxies. We will start by revisiting some of the  radiative transfer that is needed to understand how physical properties are derived from observations of \HI\ in emission and absorption.  Our vision is that this article can simultaneously serve as an overview for the new researcher on the Milky Way and external galaxies, demonstrate the constant evolution of the field, and convey the excitement of what new telescopes and simulations will reveal about the Milky Way atomic hydrogen in the coming decade.  For extragalactic researchers this review should complement the review of the Cold Interstellar Medium of Galaxies in the Local Universe by \citet{saintonge22}.

\section{FUNDAMENTALS}
Neutral hydrogen, as traced by the hyperfine transition at 21-cm, is usually described by its spin or excitation temperature, $T_s$ which is related to the H number density of two states of the hyper-fine splitting, $n_2$ and $n_1$, by the 
Boltzmann equation  as 
\begin{equation}
	\frac{n_2}{n_1} = \frac{g_2}{g_1} e^{-h \nu /k T_s}, 
\end{equation}
where $g_2=3$ and $g_1=1$ are the statistical weights for the \HI\ line.  The spin temperature is not necessarily the same as the kinetic temperature, $T_k$ of the gas. Instead, $T_s$ is determined by the Lyman $\alpha$ (Ly-$\alpha$) photon field, the pervasive radiation field around the rest wavelength of the hyperfine transition (such as the CMB), and the kinetic temperature and density, which set the collisional excitation rate \citep{field58,liszt01}.  Connecting $T_s$ and $T_k$ requires detailed understanding of {\it all} excitation processes for the 21-cm line.  In the high-density CNM, the 21-cm transition is thermalized by collisions with electrons, ions, and
other \HI\ atoms, which drives $T_s$ towards $T_k$.  In general, observers assume that $T_s \approx T_k$ for most observations of temperatures up to $\sim 1000$ K.

At the low densities typical for the WNM, collisions cannot thermalize the 21-cm transition and therefore $T_s < T_k$ \citep{field58, wouthuysen52,deguchi85, kulkarni88,liszt01}.  The Ly-$\alpha$ radiation field can help to thermalize the transition, but a very large optical depth and many scatterings of Ly-$\alpha$ photons are required to bring the radiation field and the gas into local thermal equilibrium. While the underlying atomic physics of the thermalization process is understood \citep{wouthuysen52,field58,pritchard12}, the details of Ly-$\alpha$ radiative transfer are complicated and depend on the topology and  strength of the Ly-$\alpha$ radiation field, which are complex and poorly constrained in the multi-phase ISM.  It is still customary to assume a constant Ly-$\alpha$ radiation field, with a density of $10^{-6}$ Ly-$\alpha$ photons ${\rm cm^{-3}}$\citep{liszt01}, in theoretical and numerical calculations.   

\subsection{Overview of phases}
The balance of heating and cooling processes in the ISM results in a medium with a range of temperatures and densities.   Over a  narrow range of  pressures
\HI\ can be approximated as a two-phase medium \citep{field69,wolfire95,cox05} in which the warm and cold phases of \HI\  coexist in pressure equilibrium.   Of course, the ISM is not entirely atomic; \citet{mckee77} proposed the three-phase model in which CNM clouds  are surrounded by the WNM and those neutral clouds are embedded within a supernova heated hot, ionized medium \citep{cox74}  - all in pressure equilibrium set by the supernovae.   Subsequent refinements to the theoretical models noting, for example, the effects of turbulent pressure 
have led to a much more complex structure of phases in the ISM and the understanding that the CNM plus WNM can occupy as much volume as the hot, ionized medium \citep[e.g.][]{wolfire03,bialy19,audit05,hill18} . 

\begin{marginnote}
	\entry{CNM}{Cold Neutral Medium; $T_s = 25 - 250$ K, 28-40\% of the observed \HI\ mass }
	\entry{UNM}{Unstable Neutral Medium;  $T_s = \sim250 - 4000$ K, $\sim 20 - 28$\% of the observed \HI\ mass}
	\entry{WNM}{Warm Neutral Medium; $T_s = 4000 - 8000$ K, $\sim 32 - 64$ \% of the observed \HI\ mass}
\end{marginnote}

\begin{figure}[h]
	\centering
	\includegraphics[width=\textwidth]{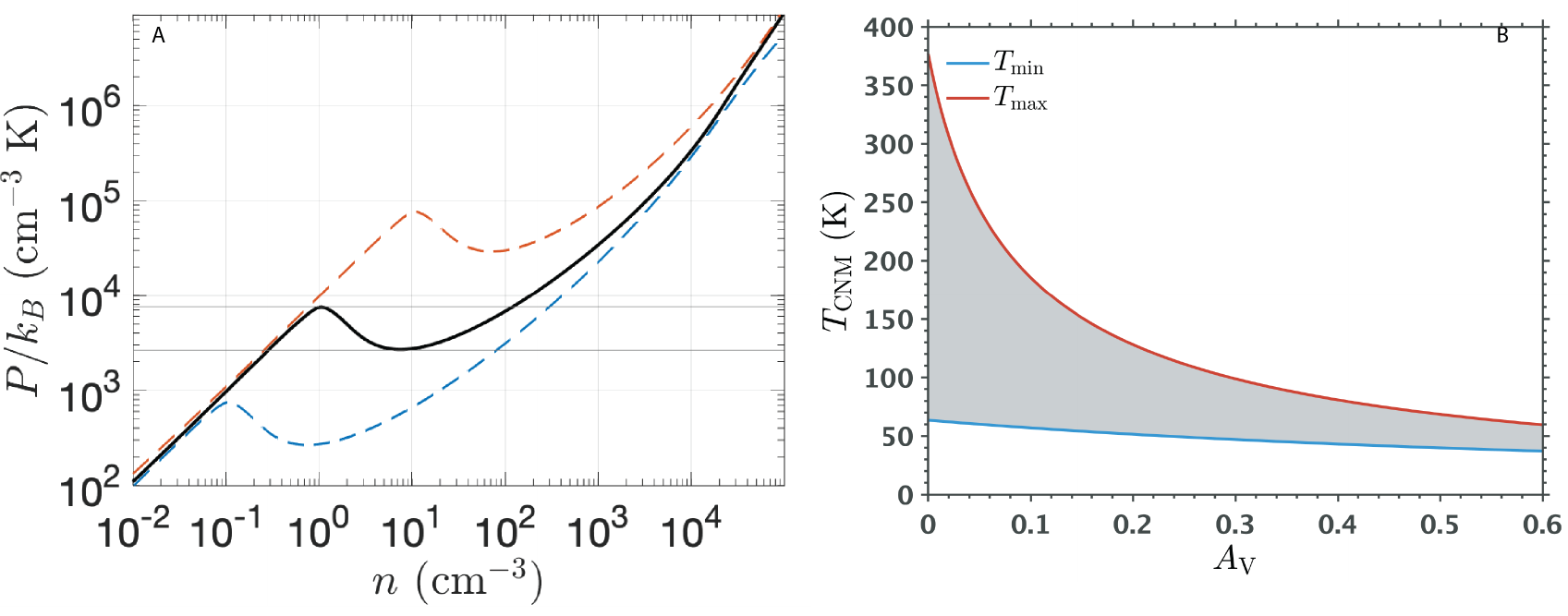}
	\caption{(a): Phase diagram for heating and cooling balance in the ISM.  The solid black line is for solar neighborhood metallicity and  UV radiation field. The orange and blue curves are for UV radiation fields that are 0.1 and 10 times the Solar neighborhood levels.  For the multi-phase \HI\ to exist, thermal pressure is required to be within a well-defined range, $P_{\rm min}$ to $P_{\rm max}$ denoted by the horizontal lines.    Reproduced with permission from \citet{bialy19}.  (b): The expected CNM kinetic temperature based on the heating-cooling equilibrium \citep{bialy19,bialy20}. At higher optical extinction, \HI\ is shielded from external radiation by dust, which  results in a lower radiation field and reduces the range of CNM temperatures allowed. Figures provided by S.\ Bialy.}
	\label{f:shmuel-plot}
\end{figure}

The  thermal equilibrium  range of pressures where the WNM and CNM can both exist is set by the balance of heating, $\Gamma$, and cooling, $\Lambda$, leading to the well-known phase diagram showing thermal pressure as a function of density 
($ P = T \times \Gamma/\Lambda(T)$).   
As shown in Figure~\ref{f:shmuel-plot}a (black line), for solar metallicity, the range of stable pressures where both WNM and CNM can exist is $1500< P_k < 10^4~{\rm cm^{-3}~K}$, with the upper limit typically closer to $\sim 3000~{\rm cm^{-3}~K}$ \citep{field69,wolfire03}.  Along the curve where $dP/dn>0$ gas is thermally stable, meaning that if a gas parcel is perturbed from the equilibrium, it will either cool or heat to return to the same equilibrium position.  These positions define the WNM at the low density end and  the CNM at the high density end.   There is another equilibrium point on the phase diagram where $dP/dn<0$ that corresponds to the intermediate temperature and density regime.  This region is thermally unstable, meaning that if a \HI\ parcel is perturbed out of equilibrium it will either cool or heat such that it moves to one of the two stable solutions.  This regime defines the unstable neutral medium (UNM).
Based on theoretical and numerical studies, the UNM temperature is expected to be in the range $\sim250-4000$ K, although this depends on the details of the local heating and coolin  \citep{field69,wolfire95a,wolfire03}.

The dominant heating mechanism for diffuse ISM is the photoelectric effect on dust grains. 
While the heating rate per unit volume depends on the starlight intensity and is a function of grain properties, including charging (which depends on the gas ionization), it does not depend strongly on temperature. As discussed by \cite{bialy19}, at very high densities or very low metallicities H$_2$ heating and cooling becomes important.   On the other side of the balance equation, the key cooling mechanism for cool \HI\ (10--10$^4$ K) is the [CII] 158 $\mu$m fine structure line emission, followed by the [OI] 63 $\mu$m fine structure line emission. As these transitions are excited by collisions with H atoms and electrons, the cooling rate depends on the fractional ionisation.  At warmer temperatures ($\gtrsim10^4$ K)  Ly-$\alpha$ emission is the dominant coolant.  As these are collisional processes, the cooling rate per volume depends strongly on temperature, $\Lambda(T)$.

The details of the classical steady-state heating and cooling processes and the resulting \HI\ phases were revised in recent modern treatments by \cite{wolfire03} and \citet{bialy19}.     \cite{bialy19} explored a broad parameter space of metallicity, interstellar radiation field, and the cosmic-rate ionization rate and showed how these effect the thermal structure of \HI.  As local conditions of the ISM change, so do the \HI\ phases.
For example, at low metallicity (0.1 Solar) and the case of dust and metal abundance being reduced by the same amount, the photoelectric heating is less effective resulting in the  expectation that the CNM should be colder.  As shown in Figure~\ref{f:shmuel-plot}a (orange curve), when the UV radiation field increases by a factor of 10, the equilibrium curve shifts to the right and the CNM and WNM  co-exist at a higher pressure.

Other local conditions, such as  column density, can also modify the \HI\ thermal structure. Figure~\ref{f:shmuel-plot}(b) shows how varying the column density or optical extinction $A_V$ has a strong effect on the range of predicted CNM temperatures, $T_{\rm min}$ to $T_{\rm max}$.  
At low $A_V$ the CNM temperature has a broad distribution ($\sim 70-350$ K), while at $A_V>0.2$ the CNM temperature range becomes very narrow, settling around 50 K for  $A_V \sim 1$.  This is due to dust shielding effectively reducing the ambient radiation field. 


Most of the detailed analytical prescriptions for heating and cooling processes consider  steady-state thermal equilibrium, e.g. \cite{mckee77,wolfire03,bialy19}. 
However, supernovae, stellar winds and many other dynamical processes constantly disrupt the heating-cooling balance and broaden the pressure distribution function.  Furthermore, turbulent processes constantly drive gas into the unstable gas regime.  To observe how various dynamical sources set the CNM/UNM/WNM distributions we require numerical simulations.

\subsection{Observable quantities}
\label{subsec:obs_quantities}
The \HI\ line is observed in either emission or absorption, depending on the optical depth $\tau (v)$ and, by extension, the temperature of the gas, $T_s$. 
Based on the physics of spectral line profiles (and assuming a Gaussian line profile),
the peak optical depth of the \HI\ line 
is related to the gas spin temperature, the (FWHM) linewidth, $\delta v$, and the column density, $N_{HI}$ as: 
\begin{equation}
	\tau_0 = \frac{N_{HI}~[{\rm cm^{-2}}]}{1.94 \times 10^{18}~[{\rm cm^{-2}~K^{-1}~km^{-1}~s}\,]} \,  \frac{1}{T_s\, [{\rm K}] \, \delta v \,[{\rm km~s^{-1}}]}.
\end{equation}
For a given column density of \HI, $\tau_0$ is inversely proportional $T_s$.
In fact, because $\delta v$ for a Boltzmann thermal distribution goes as $\delta v= 0.213 \sqrt{T_k}~{\rm km~s^{-1}}$, when $T_s\approx T_k$ 
then $\tau_0 \propto T_s^{-3/2}$ . If we assume a  WNM temperature of $T_s \sim 8000~{\rm K}$, which has a thermal velocity linewidth of $\delta v \sim 19~{\rm km~s^{-1}}$, we can easily see that for any column density observed in the Milky Way ($N_{HI} <10^{22}~{\rm cm^{-2}}$) the WNM is optically thin ($\tau_0<10^{-2}$) and would require sensitive observations to be detected in absorption.  

A primary observational property of \HI\ in emission is the brightness temperature, $T_b(v)$, which is measured as a function of Doppler velocity and related to the optical depth, $\tau(v)$, and the spin temperature, $T_s$,  as:
\begin{equation}
	T_b(v) = T_s (1- e^{-\tau(v)}).
	\label{eq:Tb}
\end{equation}
Again considering WNM temperatures of 8000 K and if $\tau\ll 1$ the brightness temperatures are approximated as $T_b \approx T_s \tau$ and can range from very small to $\sim 100$ K.  Similar arguments about temperature and density show that  \HI\ with non-negligible optical depths,  observed via absorption, is mostly CNM.    For most places in the Galaxy, gas is only atomic if it has $T_s > 25$ K, implying a thermal linewidth of $\delta v \sim 1~{\rm km~s^{-1}}$ and so $\tau > 1$ for column densities as low as $1 \times 10^{21}~{\rm cm^{-2}}$.  

It can be tempting to assume that \HI\ absorption mostly traces CNM and that \HI\ emission traces WNM, but the definitions of phases in emission is not nearly so clear.   If $T_s\sim 50$ K, optical depths of $\tau \approx 0.5$ still produce \HI\ brightness temperatures that are significantly larger than the  Kelvin or sub-Kelvin observational limits.  Conversely,   observations with very high optical depth sensitivity can detect the Warm, or at least the Unstable, Neutral Medium in absorption \citep{roy13a,murray18}.    The separation between the WNM and CNM in both emission and absorption therefore relies on high sensitivity observations and high spectral resolution for resolving narrow linewidths.

To calculate the column density of a {\em single} homogeneous \HI\ feature we need its  excitation temperature and the optical depth:
\begin{equation}
	N_{HI} = C \, T_s \int \tau(v) \, dv,
	\label{eq:column_density}
\end{equation}
where $C = 1.823 \times 10^{18}~[{\rm cm^{-2} (km~s^{-1}~K)^{-1}}]$. If there are several \HI\ features along the line of sight, the column density for each feature must be treated individually with its own $T_s$ \citep[e.g.][]{heiles03a}.  In the absence of full information about the distribution of spin temperatures with respect to velocity, it is often necessary and prudent to assume that the gas is isothermal, such that only one temperature is represented at each velocity channel, where
\begin{equation}
T_s(v) \approx \frac{T_b(v)}{1-e^{-\tau(v)}}.  
\end{equation}
The column density of the entire sight-line can then be written
\begin{equation}
N_{HI,\rm{iso}} = C \int \frac{\tau(v) T_b(v)} {(1-e^{-\tau(v)})}dv,
\label{eq:NH_iso}
\end{equation} as proposed by \citet{dickey82} and developed by \cite{chengalur13}.
Clearly, for small optical depths Equation \ref{eq:NH_iso} reduces to the well known simplification
\begin{equation}
	N_{HI,\rm{thin}} = C \, \int T_b(v) dv.
	\label{eq:NH_thin}
\end{equation}

As described above, \HI\ in emission can trace both the CNM and the WNM.  \HI\ emission on its own provides an estimate of total column density if we assume optically thin emission ($\tau\ll 1$, Equation \ref{eq:NH_thin}), and an upper limit on the kinetic temperature of the gas, $T_k$ through its FWHM linewidth, $\delta v$:
 \begin{equation}
 T_k[K] \leq 22 \, (\delta v[{\rm km~s^{-1}}])^2.	
 \end{equation}
The CNM temperatures of $\sim 20$ -- $200$ K produce thermal linewidths of only $1$ -- $2~{\rm km~s^{-1}}$, while WNM temperatures of $\sim 3000$ -- $8000$ K  produce thermal linewidths of  $11$-$20~{\rm km~s^{-1}}$. 

\HI\ observers directly measure optical depth $\tau(v)$ through absorption, and brightness temperature, $T_b(v)$ through emission, plus the linewidth, $\delta v$, of individual features in the spectrum.  However, 
\HI\ emission and absorption can be combined to solve for the spin temperature, $T_s$ of an \HI\ feature, as described in Equation~\ref{eq:Tb}.  
From here, and by using the linewidth from \HI\ emission, the contributions from turbulent motions of the gas can be estimated \citep{heiles03b}.

\section{NUMERICAL SIMULATIONS OF \HI}
Numerical simulations of the ISM are invaluable for understanding how the many, varied physical processes of the ISM shape \HI\ over a wide range of environments. 
Huge advances in computational facilities and techniques in recent years have 
moved from 2-D idealized boxes with basic feedback processes  driven artificially at a fixed rate \citep{audit05} to 3-D, self-consistent, parsec-scale, multi-phase simulations that track the effects of  different feedback sources, e.g.\ SNe, stellar winds, cosmic rays, and radiation \citep{kim13,kim17,rathjen21}.
Even still, numerical simulations are not able to provide the dynamic range needed to self-consistently track the \HI\ from large-scale infalling, diffuse gas all the way to molecular clouds within a larger galactic context (outflows, disk-halo interface regions, galactic rotation).  
However, many simulations have demonstrated the importance of including the large-scale fundamental processes to shape \HI\ in realistic ways.  Hopkins et al.\ (2012, 2018, 2020)\nocite{hopkins12,hopkins18a,hopkins20}, show that stellar feedback is essential to reproduce realistic \HI\ morphology observed in the Milky Way and other galaxies, without it simulated gas density distributions are too high and lead to runway star formation and unrealistic phase distributions.  Similarly, the distribution of spatial power, represented by the spatial power spectrum, requires stellar feedback to match observations \citep{iffrig17,grisdale18}.

Over the last two decades many simulations have focused on high resolution models of the highly dynamic and turbulent character of the ISM trying to reproduce the observed thermal and morphological structure of \HI. For example, \citet{audit05} showed that a collision of incoming turbulent flows can initiate fast condensation from WNM to CNM. 
\citet{koyama02} and \citet{maclow05} explored how shocks driven into warm, magnetized, and turbulent  gas by supernova explosions create dense, cold clouds.  The latter study showed a continuum of gas temperatures, with the fraction of the thermally-unstable \HI\ constrained by the star formation rate. \cite{ntormousi11} modeled the formation of cold clouds in a realistic environment of colliding superbubbles, and showed that cold, dense and filamentary structures form naturally in the collision zone. 
As \HI\ is the seed for the formation of H$_2$ many simulations start with initial conditions for the diffuse \HI\ and follow development of the CNM first, and then H$_2$ \citep{glover10,clark12,valivia16,vazquez-semadeni10,seifried22}.
Simulations such as \citet{vazquez-semadeni10} and \citet{goldbaum11} have suggested that
the CNM plays a central role in the formation and
evolution of molecular clouds via accretion flows.
However,  exactly how the \HI\ cycles through various phases in the process of building molecular clouds is still not clear. Numerical simulations by \cite{dobbs12} suggested that it is the UNM that is direct fuel for molecular clouds.

\begin{figure*}[h]
	\includegraphics[scale=0.68]{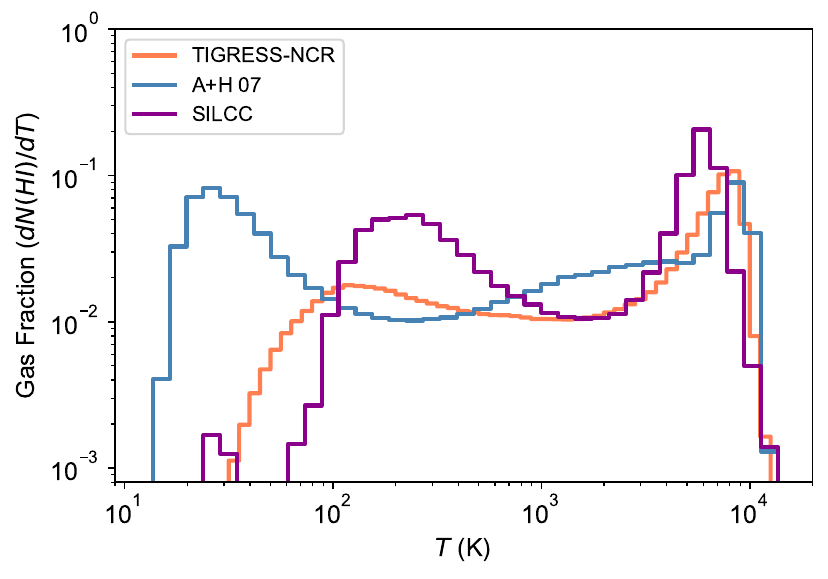}
	\caption{Fraction of total column density at a given spin temperature as predicted from the 3-D simulations TIGRESS-NCR \citep{kim22} and SILCC \citep{rathjen21} and the 2-D simulation by Audit \& Hennebelle (2007, labeled A+H 07). Data were kindly provided to the authors by C.-G.\ Kim, T.-E.\ Rathjen \& P. Hennebelle. 
	}
	\label{fig:Ts_TIGRESS}
\end{figure*}

A key power of numerical simulations is in providing data for a direct comparison with observations. We summarize two topics where such comparisons are important: \HI\ mass fraction and synthetic line profiles.  

\subsection{The \HI\ mass fraction as a function of phase}
The mass fraction of different \HI\ phases is a useful quantity to predict because it can be constrained observationally, which helps to clarify the processes of importance.
The existence of the bi-modal \HI\ distributions has become a stable feature in recent simulations.  While the inclusion of different feedback sources (supernovae, stellar winds, UV radiation from HII regions) initially resulted in a broad density distribution without two well-separated \HI\ phases \citep[e.g.,][]{maclow05}, more recent magneto-hydrodynamic simulations easily produce the bi-stable \HI\ structure with the CNM and WNM as distinct density peaks \citep{kim14,kim17,hill18}.  Two suites of simulations with the highest resolution that focus on the thermal state of \HI\ are TIGRESS \citep{kim17} and SILCC \citep{walch15,rathjen21}.    The TIGRESS simulations show that although turbulence, temporal fluctuations of the heating rate, and expanding superbubbles continuously populate the UNM phase, thermal instability and rapid cooling reduce the amount of UNM and result in the majority of the \HI\ following the thermal equilibrium and being in the two-phase state (Figure~\ref{fig:Ts_TIGRESS}). Even when varying the heating rate significantly, \cite{hill18} showed that the two-phase \HI\ persisted and the mass fraction of the thermally unstable \HI\ was $<20$\%. 

While recent simulations agree about the existence of the bi-stable \HI, significant differences exist in the mass fraction across different phases.
Figure~\ref{fig:Ts_TIGRESS} shows predictions for the \HI\ fraction of total column density at a given $T_s$ from several simulations. SILCC predicts a significantly higher CNM mass fraction ($\sim50$\%) relative to TIGRESS ($<20$\%). The UNM fractions seem to agree at  $\sim20-30$\%.  
Surprisingly, different feedback channels do not have a large effect on the \HI\ phase fractions in TIGRESS (Kim, private communication), while slightly larger differences are seen in SILCC \citep{rathjen21}. 

\subsection{Synthetic \HI\ line profiles}

Synthetic \HI\ spectra in emission and absorption are essential for comparing observations with simulations. Through direct comparison it is possible to test the reliability of simulations and investigate biases in observational processing techniques \citep{haworth18}. As an example, using TIGRESS \cite{kim14} provided thousands of synthetic spectra that \cite{murray18} analyzed exactly as observed \HI\ spectra. 
By comparing identified spectral features in the synthetic spectra with 3D density structures, it was possible to estimate the completeness and accuracy of the radiative transfer approach used to observationally estimate  $T_s$. Similarly, considering correspondence between the true positions and observed radial velocities 
of molecular clouds, \citet{beaumont13} showed that the superposition of clouds 
along the line of sight introduces significant uncertainty to observational estimates of cloud mass, size and velocity dispersion.
While this study specifically focused on CO and denser environments, masses and sizes of \HI\ structures are similarly affected by the line-of-sight complexities. Simulated spectra will become even more important in the future  to assess the accuracy of automated data handling routines that will be used by large surveys.

The comparison of simulated and observed spectra is important to fully constrain the excitation processes for the WNM.
A large uncertainty exists in excitation temperature of \HI\ ($T_s$) especially for the WNM where the collisional excitation is insignificant. Excitation by Ly-$\alpha$ resonant scattering can be a dominant excitation mechanism \citep{liszt01,kim14}. Recently, direct Ly-$\alpha$ radiation transfer in a realistic ISM reveals that the Ly-$\alpha$ excitation in the solar vicinity is efficient enough to make $T_s$ as high as gas kinetic temperature $T_k\sim7000-8000$ K \citep{seon20}, in agreement with the observational result of \citet{murray14}.


\section{OBSERVATIONS OF \HI}
Observations of \HI\ have overwhelmingly focused on \HI\ emission, which is used to study the structure of \HI, measure column density with assumptions about optical depth, and measure velocity linewidths.  Recently, \HI\ absorption surveys have increased in both coverage and sensitivity.  However, absorption observations are targeted towards a limited set of directions but provide measurements of optical depth as well as linewidth.  The power of \HI\ is really unlocked when absorption and emission are brought together.

\subsection{\HI\ in Emission}\label{subsec:HIemission}

Based on the relations in Section \ref{subsec:obs_quantities},
the brightness temperature of \HI\ emission is proportional to column density, 
resulting in the \HI\ emission being
a powerful tracer of the distribution of \HI. Furthermore, it is possible for modern observational surveys to measure \HI\ emission at all spatial positions, producing a fully-sampled atlas of the distribution of \HI.  Historically, this emission has been used to describe the 3-D distribution of the Milky Way 
(Section \ref{subsec:3d_distribution}), as well as the overall morphology of the atomic ISM 
(Section \ref{sec:structure}). Recently, the spatial structure of \HI\ emission has been extended to trace the disk-halo interaction (Section \ref{sec:structure})
and to show its connection with magnetic fields 
(Section \ref{subsubsec:magnetism}).  Relying on its complete sampling of all spatial scales, statistical studies of \HI\ emission have been used to elucidate  the turbulent properties of the medium (see Section \ref{subsubsec:SPS}).

For many years \HI\ observers have been working to decompose the WNM and CNM components of the emission on the basis of velocity dispersion \citep[e.g.][]{kulkarni85,verschuur95}.  The decomposition is challenging because the WNM is bright, spatially pervasive and has broad-linewidths, whereas the CNM has narrow-linewidths and is weak unless $\tau \approx 1$, meaning that its contribution to $T_b(v)$ is easily swamped by the WNM. 
Recently, sophisticated techniques for semi-autonomous  decomposition have been  used on large-area \HI\ emission surveys to derive estimates of the gas phase fractions of CNM, UNM and WNM and to consider differences in the structure of the WNM and CNM \citep{haud07,kalberla18, marchal19}.  For example, \citet{kalberla18} used Gaussian decomposition on the HI4PI survey \citep{hi4pi-collaboration16} to show that the CNM is largely filamentary.   \citet{marchal21b} used a different decomposition technique called ROHSA \citep{marchal19} that makes use of spatial continuity between adjacent  spectra to derive a regularized Gaussian fit over a high-latitude \HI\ field.  \citet{murray20} took a different approach and used machine learning with a Convolutional Neutral Network,  informed by \HI\ absorption spectra, to separate the CNM from WNM in emission.  These varied techniques provide estimates of CNM fraction and show structural differences between the phases.

\subsubsection{Summary of 21-cm Emission Surveys}
\HI\ surveys of emission in the Milky Way have traditionally been carried out with single dish radio telescopes.    With each decade since the discovery of the \HI\ line in 1951 by \citet{ewen51} the angular resolution and sky coverage of \HI\ surveys has increased by roughly a factor of two.  Last century ended with the famous Leiden-Argentine-Bonn (LAB) compilation \citep{kalberla05} providing a uniform survey at 30\arcmin\ angular resolution, $1~{\rm km~s^{-1}}$ velocity resolution and $\sim 100$ mK brightness sensitivity over the whole sky.  As reviewed in \citet{kalberla09}, the LAB survey formed the basis for much of what we know about the large-scale \HI\ structure of the Milky Way.  

At the end of the 20th century two technical advances, multi-beam receivers and interferometric mosaicing, came to radio telescopes and pushed \HI\ surveys into a new era.  The first of these, the multi-beam receiver, decreased the survey time required to reach high sensitivity over extensive sky areas with large ($60$ - $300$ m) telescopes.  The multi-beam receivers on the Parkes (13 beams), Effelsberg (7 beams), and Arecibo (7 beams) telescopes all produced a new generation of surveys covering large areas at resolutions less than 16\arcmin.  The Parkes Galactic All-Sky Survey \citep[GASS;][]{mcgriff09,kalberla10,kalberla15} and Effelsberg-Bonn \HI\ survey \citep[EBHIS;][]{winkel10}  were combined together to produce HI4PI \citep{hi4pi-collaboration16} at 16\arcmin.  The all-sky composite HI4PI shows structures to be traced across the whole sky and enables statistically complete studies of phase distributions to be conducted in different regions with the same data. The most recent, and highest resolution, of the single dish surveys is GALFA-HI  conducted with Arecibo \citep{peek18}.  GALFA-HI has pushed single dish  \HI\ emission surveys into a new realm of sensitivity coupled with angular and spectral resolution.  Unlike many of the previous surveys of \HI\ emission, GALFA-HI has a velocity resolution of $\sim 0.2$ \kms, revealing very fine velocity gradients across the \HI\ emission sky.  
%

Simultaneously with multi-beam receivers, interferometric mosaicing came into common practice.  Interferometric mosaicing, which combines observations of many overlapping fields-of-view, enabled observations of a large area of sky {\em and} the recovery of  larger angular scales than accessible through a single pointing.  Mosaicing was originally proposed by \citet{ekers79}, but was not widely adopted for diffuse imaging until the 1990's when \citet{sault94} perfected the technique for the Australia Telescope Compact Array (ATCA) starting with  the Small  Magellanic Cloud \citep{staveley-smith97}.  The next development, combining interferometric mosaics with image cubes from single dish telescopes \citep{stanimirovic99, stanimirovic02a} resulted \HI\ cubes that recovered emission on all angular scales from many degrees to  the resolution limit of the array, typically 1 - 3 arcminutes.  The technique was adopted by the Dominion Radio Astrophysical Observatory's Synthesis Telescope with the Canadian Galactic Plane Survey \citep[CGPS;][]{taylor03} and the DRAO \HI\ Intermediate Galactic Latitude Survey \citep{blagrave17}, the ATCA's Southern Galactic Plane Survey \citep[SGPS;][]{mcgriff05} and, finally, the Very Large Array (VLA)'s Galactic Plane Survey \citep[VGPS;][]{stil06}.  

The time required to reach a given brightness sensitivity goes as the inverse square of the longest baseline of the interferometer forcing a constant trade-off between resolution and brightness sensitivity.  Because the time needed to survey large areas at high resolution and with high surface brightness sensitivity with interferometers is prohibitive, the high angular resolution interferometric surveys were limited to much smaller aerial regions than single-dish surveys, focusing primarily on the Galactic Plane and a single intermediate latitude patch.     Recently, the Very Large Array conducted a new Galactic Plane survey, THOR  \citep{beuther16,wang20}, aimed at improving the angular resolution of the VGPS to 20\arcsec\ in angular resolution.   Together, the \HI\ surveys of the last twenty years have demonstrated the inherent wealth of structure in the Galactic \HI\ over all size scales, which we will discuss in Section \ref{sec:structure} below.

\subsection{\HI\  Continuum Absorption}
The most powerful way to constrain the CNM, with its moderately high optical depths, is via absorption against continuum background sources.  To see how this proceeds let us extend the radiative transfer discussion from Section~\ref{subsec:obs_quantities}.
In general, the simple radiative transfer equation for a single \HI\ feature given in Equation~\ref{eq:Tb} will include a contribution of the diffuse radio continuum emission temperature of the sky, $T_{c,sky}$ that incorporates
the CMB and the Galactic synchrotron emission, varying with position.   
Given a bright background radio source with a continuum spectrum, $T_{c, bg}$, (usually a quasar or a radio galaxy) absorption measurements can be made on-source.  The radiative transfer equation in the direction of the background source then becomes:
\begin{equation}
	T_b(v) = T_s(1 - e^{ -\tau(v)}) + (T_{c,sky} + T_{c,bg}) e^{-\tau(v)}.
	\label{eq:simple_RTE_on}
\end{equation}
Comparing \HI\ spectra obtained in the direction of the background source and spectra close to the background source  (``off-source", where $T_{c,bg}=0$ K), under the assumption of uniform foreground \HI\ emission, we can solve Equation ~\ref{eq:simple_RTE_on} for both $T_s$ and $\tau(v)$. 
Equation~\ref{eq:column_density} then provides an estimate of the column density, $N_{H\textsc{i}}$ of \HI\ structures along the line of sight.

For the uniform \HI\ emission assumption to work, high angular resolution is required so that the off-source spectrum samples a very similar sight-line to the on-source spectrum. In this respect, interferometers are highly advantageous and they resolve out the large-scale structure to help simplify Equation~\ref{eq:simple_RTE_on} with $T_{c,sky}\approx 0$ K. 
Because interferometric mapping with the inclusion of all spatial scales is time consuming, single-dish telescopes are often used for the off-source measurements of $T_b(v)$. However, as the emission fluctuations within the single-dish beam can
contaminate absorption measurements, more complex strategies for estimating spatial variations of \HI\ emission are required \citep[e.g.][]{heiles03a}.  A complicating factor when using single-dish telescopes to provide off-source spectra is  the mismatch of beam sizes 
of emission and absorption spectra when solving the radiative transfer equations for $T_s$ and $\tau(v)$. 
The \HI\ absorption spectrum samples a very small solid angle occupied by the background (often point) source, while the \HI\ emission is typically measured on arcminutes scales (Section 4.1). This can result in absorption features not having clear corresponding emission components. 

The biggest complication in deriving both $T_s$ and $\tau(v)$ is that both 
$T_b(v)$ and $\tau(v)$
spectra usually contain  multiple velocity components, resulting in the need for more sophisticated radiative transfer calculations.
Several different approaches have been used to handle this, e.g.\ fitting individual components with Gaussian functions \citep[e.g.,][]{heiles03a}, the slope method \citep[e.g.][]{mebold97,dickey03} or simply working with integrated spectral quantities (e.g.\ Equations 5 and 6). This problem is exacerbated at low Galactic latitudes where lines of sight  have many velocity components.  A related concern is whether selected velocity structures correspond to real physical structures in the ISM or are possibly seen as superpositions of many unresolved structures. Such superpositions introduce biases in observational estimates of $T_s$ and $\tau(v)$ and have been investigated by numerical simulations \citep[e.g.,][]{hennebelle07,kim14}.  \citet{murray17} 
used synthetic spectra from \citet{kim14} and showed that the completeness of  recovering $T_s$ with Gaussian fits depends strongly on the complexity of \HI\ spectra. For simulated high-latitude lines-of-sight ($|b| > 50$ deg)
99\% of identified structures in radial velocity had corresponding density features.
However, the recovery completeness dropped  to 
67\% for 20 $< |b| < 50$ deg and 53\% for 0 $< |b| < 20$ deg. As  line-of-sight complexity increases, completeness 
decreases,
reflecting the difficulty
in unambiguously associating  spectral features in  emission and absorption in the
presence of line blending and turbulence.

\subsubsection{Summary of 21-cm Continuum Absorption Surveys}

\begin{figure}[h]
	\includegraphics[width=\textwidth]{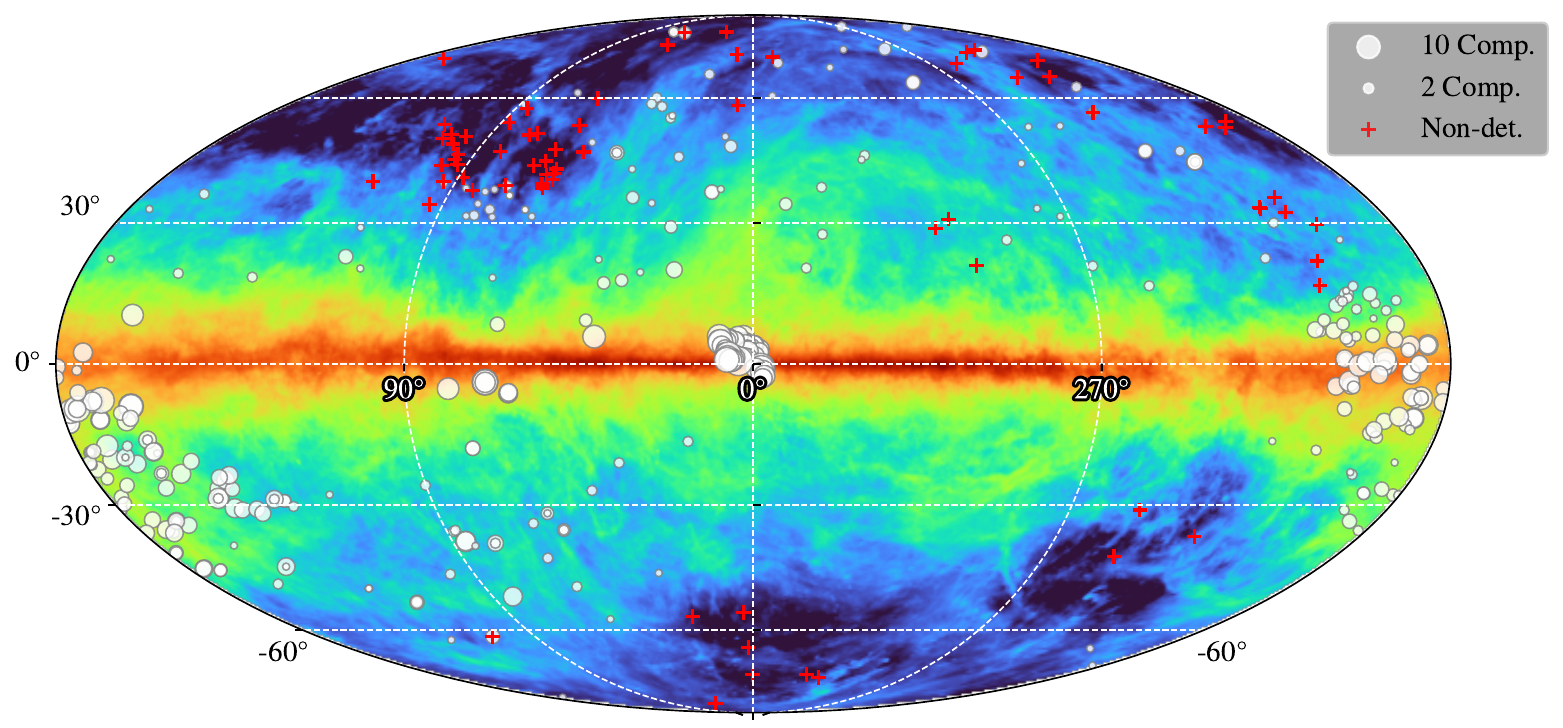}
	\caption{Compilation of \HI\ absorption spectra from the \bighicat\ catalog overlaid on a map of \HI{} emission integrated from -75~\kms{} to +75~\kms{} \citep{hi4pi-collaboration16}. Non-detections are shown as crosses; detections are sized according to the number of absorption components. Coordinates are Galactic.}
	\label{f:abs-positions}
\end{figure}

Due to the high optical depth of the CNM, 21-cm absorption signatures have been easy to detect, even with low sensitivity observations 
\citep[e.g.][]{hughes71,knapp72,radhakrishnan72,lazareff75,dickey77,dickey78,payne78,crovisier78a,dickey82,belfort84,braun92,heiles03a}. Perhaps the most influential single-dish absorption survey was the Millennium Arecibo 21 Centimeter Absorption-Line Survey,  comprised of 79 \HI\ absorption and emission spectral pairs spread over the full Arecibo Observatory sky \citep{heiles03a}.

The optical depth of the UNM and WNM is very low ($\tau\leq10^{-2.5}$) and 
highly sensitive observations 
are required to detect these phases in absorption. A few studies have targeted individual detections of the WNM in absorption to directly measure $T_s$ \citep[e.g.,][]{carilli98,dwarakanath02,murray14}, while many other surveys estimated WNM $T_s$ from upper limits (i.e.\ $T_k$) or as line-of-sight averages in the presence of strongly absorbing CNM gas \citep{mebold82,heiles03a,kanekar03,kanekar11,roy13b}. 
To clearly detect broad and weak absorption lines from the WNM requires excellent spectral baselines and is best done using interferometers.  The upgrade of the Karl G.\ Jansky Very Large Array (JVLA), resulted in 
the bandpass stability
high enough to detect shallow ($\tau_0 \sim10^{-3}$), wide ($\delta v \sim7-8\rm\,km\,s^{-1}$) absorption lines \citep{begum10}. In combination with \HI\ emission from the Arecibo Observatory, Begum et al.\ identified individual absorption lines in the UNM regime with $T_s=400-900\rm\,K$.   Roy et al.\ (2013a,b)\nocite{roy13a,roy13b} used the Westerbork Synthesis Radio Telescope (WSRT), Giant Metrewave Radio Telescope (GMRT) and the  ATCA 
for another deep \HI\ absorption survey detecting the UNM. These studies emphasized the need for larger samples of interferometric detections of \HI\ absorption at high sensitivity to  constrain the fractions of gas in all \HI\ phases. 

The 21-SPONGE project \citep{murray14,murray15,murray18} was the one of the most sensitive absorption line surveys. It targeted 58 bright background sources 
reaching sensitivity of  $\sigma_{\tau}<1 \times 10^{-3}$ 
to detect in absorption \HI\ in all stable and thermally unstable phases. 
The \HI\ absorption spectra were complemented with  \HI\ emission from Arecibo, and a streamlined, reproducible  fitting and radiative transfer approach used to constrain physical properties of \HI. 

While targeted \HI\ absorption surveys are time consuming,  large interferometric surveys of the Milky Way plane (CGPS, SGPS, VGPS, THOR) have had high enough spatial resolution to allow extraction of both \HI\ absorption and emission spectra in the direction of background sources \citep{strasser07,dickey03,dickey09}. For example, this approach enabled a large sample of $\sim300$ absorption-emission pairs that was used in \cite{dickey09}.
While providing important constraints about the spatial distribution of the CNM, these interferometric surveys mainly used integrated properties as \HI\ spectra within $\pm$ few degrees from the plane are very complex. 

Recent 21-cm \HI\ absorption and emission surveys are summarized in Table 1 
of the Supplementary Material. From a subset of these we have compiled a catalog containing key observed properties of \HI\ absorption in the Milky Way.  The compilation, which we call \bighicat{} and describe in the Supplementary material, combines publicly available spectral Gaussian decompositions of several 
surveys. In total, \bighicat{} comprises 372 unique lines of sight and 1223 Gaussian absorption components giving $T_s$, peak $\tau$ and other properties. Figure \ref{f:abs-positions} shows the distribution of these lines of sight.

\subsection{\HI\  Self-Absorption}
Another manifestation  of \HI\ absorption is \HI\ self-absorption (HISA), which occurs when cold \HI\ lies in front of warmer background \HI\ emission at the same velocity.  HISA is not, strictly speaking, self-absorption - the two \HI\ features are not co-spatial, but they are co-spectral.   \HI\  in the inner Galaxy, where velocities correspond to two distances, provides an ideal  background for detecting foreground cold \HI\ as HISA.   Broadly, HISA is evident in two situations: high optical depth environments with a moderate background, such as molecular clouds  (often referred to as HINSA, e.g. \citealt{li03,li12}) \nocite{li03,li12} or moderate optical depth CNM with a high brightness backgrounds, such as the Galactic Plane \citep[e.g.]{gibson05}.    HISA has historically been used to study the structure and temperature of cold clouds \citep[e.g.][]{heeschen55,knapp74,baker79}, but its widespread presence became particularly clear with the high-resolution afforded by the interferometric Galactic Plane surveys \citep{gibson00,gibson05a,wang20b,wang20a,beuther20}.      

\begin{figure}
	\centering
	\includegraphics[width=\textwidth]{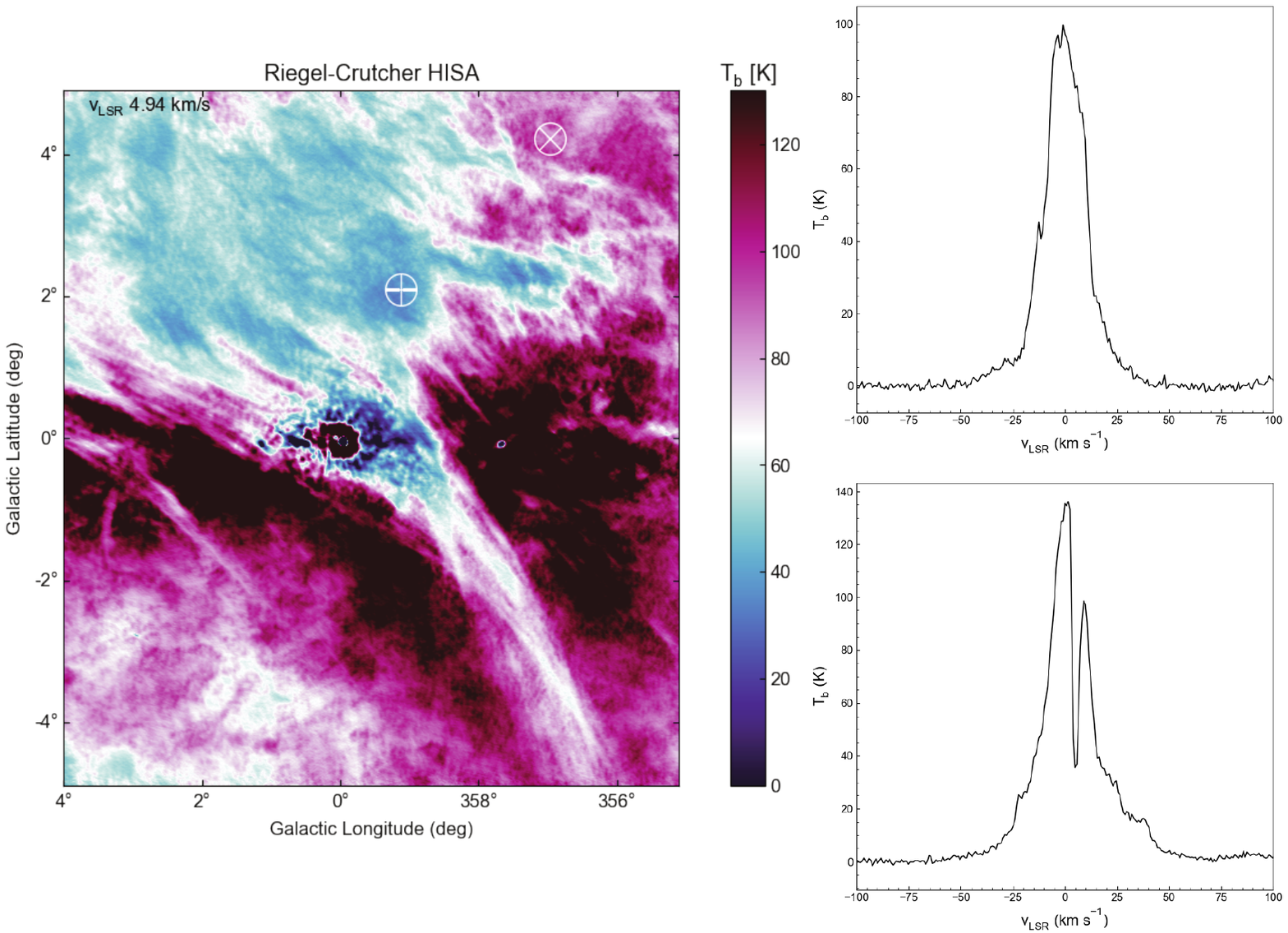}
	\caption{HISA in the Riegel-Crutcher Cloud \citep{mcgriff06b}.  The spectra on the right are from the positions marked.  The top spectrum is at the position $\otimes$ and the bottom spectrum, containing HISA, from the position marked $\oplus$. }
	\label{fig:HISA_RC}
\end{figure}

An advantage of HISA over other observations of \HI\ emission is that it clearly separates CNM from WNM.  Whereas \HI\ emission is usually dominated by the WNM, and  the physical properties 
of the CNM are primarily studied in discrete lines-of-sight towards continuum sources (an HI-absorption ``grid"), HISA can give continuous spatial sampling of the CNM.   In practice, however, the efficacy of spatial recovery of the cloud properties ($N_H$, $\tau$, and $T_s$) is not trivial because variations in background \HI\ emission can obfuscate the properties of the absorbing cold cloud.    

Solving the radiative transfer equation (Equation~\ref{eq:simple_RTE_on}) for HISA is even more  complicated than for continuum absorption. We have to consider not just the continuum background, $T_c$ but also the spin temperature and optical depth of all of the emitting and absorbing \HI\ along the line of sight, including the foreground ($T_{s,{\rm f}}$, $\tau_{\rm f}$) and background ($T_{s,{\rm b}}$, $\tau_{\rm b}$), and the HISA feature itself ($T_{s,{\rm HISA}}$, $\tau_{\rm HISA}$).
The expected  brightness temperature, $T_b$, observed in the direction of a cold HISA structure is therefore:
\begin{equation}
\begin{aligned}
	T_b  \, = & T_{s,{\rm f}} \,(1-e^{-\tau_{\rm f}}) \\
	  + & \, T_{s, {\rm HISA}}\,(1 - e^{ -\tau_{\rm HISA}}) \,e^{-\tau_{\rm f}} \\
	 +  &  \,T_{\rm s,{\rm b}} \,(1 - e^{ -\tau_{\rm b}}) \,e^{-(\tau_{\rm f} + \tau_{\rm HISA})} \\
	 + &  \,T_c \, e^{-(\tau_{\rm f} + \tau_{\rm HISA} +\tau_{\rm b} )}.
\end{aligned}
	\label{eq:hisa_1}
\end{equation}
All quantities except $T_c$ are functions of radial velocity.   Unlike continuum absorption where the brightness temperature without any absorption 
can be easily measured adjacent to the source, for spatially extended HISA $T_{\rm off}$ is usually estimated by interpolating over the absorption feature in velocity space.  In all HISA observations, $T_{s, {\rm HISA}}$ and $\tau_{\rm HISA}$ are degenerate.  Various approaches have been proposed to solve Equation~\ref{eq:hisa_1} for $T_{s, {\rm HISA}}$ and $\tau_{\rm HISA}$ \citep{knapp74,gibson00}, but remains a challenge.

Despite the challenges of robustly solving for $T_s$ with HISA, 
\HI\ self-absorption
provides a valuable probe of the transition zone 
between atomic and molecular gas in molecular clouds.  \citet{syed20} and \citet{wang20a}, for example, use the combination of traditional \HI\ emission, HISA-traced gas, and CO-traced H$_2$ to measure the full hydrogen column density probability density function for Giant Molecular Filaments.  The addition of HISA-traced gas allows for an optical depth correction to the \HI\ emission, giving a more realistic total column density (subject to uncertainties in assumed values for $T_s$ and the fraction of foreground to background $T_b$).   However, recent synthetic observations of SILCC-Zoom simulations have raised some concerns about the reliability of using HISA to measure $N_H$ \citep{seifried22}.  They  found that observations of HISA tend to under-estimate the column density 
by factors of as much as 3 - 10 because HISA clouds  rarely have a single temperature.  Further studies of simulations will 
help guide observers in how best to incorporate HISA in $N_H$ measurements.

Perhaps the most important, unique information that  HISA gives is the spatially resolved kinematics of the CNM.   To  spatially resolve velocity gradients in the CNM from emission observers have to separate the narrow-line CNM from the pervasive broad-linewidth WNM.  Spectral decomposition of complex \HI\ emission profiles is generally not unique \citep[see][]{marchal19}.  CNM traced by HISA, on the other hand, is relatively easy to separate spectrally from the confusing WNM, making it useful for deriving velocity fields of CNM structures.    \citet{beuther20}  used this technique to show that the HISA-traced CNM around the infrared dark cloud G28.3 is kinematically decoupled from the denser gas as traced by $^{13}$CO and [CI].  It seems certain that with increased angular resolution and sensitivity of future surveys HISA will be used more extensively.

\section{THE NATURE  OF  \HI\ IN THE MILKY WAY}
While \HI\ emission surveys have been invaluable  in revealing detailed properties of the Galaxy's gaseous structure 
\citep[see reviews by][]{kulkarni88,dickey90,kalberla09}, perhaps the greatest advances in understanding the physics of Galactic \HI\ over the last decade have come 
from studying the physical properties, overall distribution, and structure of the absorbing \HI.
The absorbing \HI{}, as we discuss in this Section, represents mainly the CNM, with a very small fraction of absorption corresponding to the UNM and WNM. 
Because of observational limitations, our understanding of the nature of the CNM of galaxies almost exclusively comes from the Milky Way.  And yet, the CNM is a critical step towards the formation of molecular clouds and therefore star formation.   While many sensitive observations of \HI\ absorption have been undertaken over the last 20 years, we are still scratching the surface on understanding CNM properties. Small samples still limit studies of the thermally unstable \HI\, and a significant debate persists about what fraction of \HI\ is in the UNM. As we move forward, the Milky Way will remain  the key place for testing details of the multi-phase physics and we anticipate great advances in this area with upcoming large surveys.

Following the first astronomical observations of absorption
and emission via the 21 cm transition of \HI\ \citep{ewen51,muller51,hagen55}, clear
differences in the observed velocity structure between 
emission and absorption spectra were attributed to significant
variations in the temperature and density of the gas along
the line of sight \citep{clark65,dickey78}.
The distinction between \HI\ structures of different phase (density and temperature) is still complex. As discussed in Section 4.2, this complexity comes from the often large overlap of different \HI\ structures in the radial velocity space, as well as the limitations posed by the observational sensitivity. 

Below we summarize what is currently known about the temperatures and mass fractions of the different \HI\ phases.  We make use of \bighicat\ to examine environmental dependencies for the CNM.

\begin{figure}
    \centering
    \includegraphics[width=\linewidth]{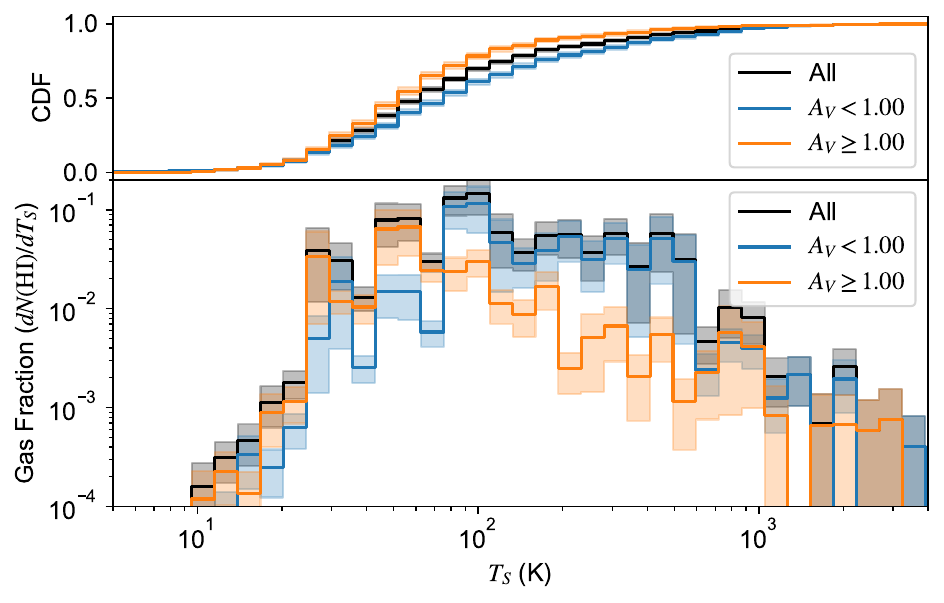}
    \caption{Fraction of total column density detected as a function of spin temperature in the  \bighicat\ \HI\ absorption compilation.  The overall fraction (black) is divided on the basis on visual extinction, $A_V$.  The median temperature is $T_s \sim 50$ K.  For  high extinction environments ($A_V>1.0 $, orange) the temperature range observed in the CNM is narrow and peaked at $T_s \approx 50$ K, whereas the lower extinction environments ($A_V<1.0$, blue) show a broader range of CNM temperatures peaked closer to $T_s \approx 100$ K. The $T_s$ cumulative distribution function (CDF) for each sample is shown on the top. Uncertainties, calculated by bootstrapping over 1000 trials, are shown as shaded regions around the different histograms. The sample is incomplete in the range $ 500~{\rm K} \lesssim T_s\lesssim 4000$ K where observational sensitivity limits the number of measurements.}
    \label{f:hi-gas-fraction}
\end{figure}

\subsection{CNM temperature} 
The CNM appears ubiquitous in the Milky Way. As seen in Figure~\ref{f:abs-positions}, it is easily detected at low Galactic latitudes, where absorption spectra often have multiple components, and high Galactic latitudes, where spectra are simple, often with a single, low optical depth CNM feature. In the \bighicat\ 306/372 directions show significant absorption in at least one component, and many \HI\ absorption surveys, including 21-SPONGE, have shown a similarly high detection rate of \HI\ absorption,  over 80\%.

In Figure~\ref{f:hi-gas-fraction} we use entries from the \bighicat\ database where individual velocity components were fitted with Gaussian functions and $T_s$ was derived to show the fraction of the total \HI\ column density as a function of $T_s$ (black line). \citet{murray18} used synthetic \HI\ spectra to demonstrate that the approach of fitting Gaussian functions to individual velocity components, in combination with radiative transfer calculations, is successful and complete at recovering the overall
fraction of the \HI\ mass that is in the CNM.  The bottom panel of Figure~\ref{f:hi-gas-fraction}  can be compared with the numerical expectations shown in Figure~\ref{fig:Ts_TIGRESS}.   The observed \HI\ gas as a fraction of total column density (or mass) peaks around 50-100 K, has a broad shoulder up to 500 K, and an extended tail up to $T_s =3000$ K.    Based on theoretical and numerical studies, the peak of the \HI\ distribution corresponds to the CNM and has been observationally straightforward to detect \citep[many references listed in Section 3.2, also][]{dickey90}.  Clearly, the (absorption) observed \HI\ gas fraction is missing the WNM portion of the expected distribution.  At this stage the decline in the gas fraction about $T_s=500$ K is an observational limitation; even with long integrations it has been very hard to detect the UNM and WNM in absorption (see Sections 5.4 \& 5.5).

The observed peak of CNM temperature distribution  of $\sim50-200$ K is largely in agreement with theoretical expectations for Solar metallicity and radiation field (Figure~\ref{fig:Ts_TIGRESS}), at low optical extinction. The  \bighicat\ catalog is biased towards higher latitude observations (due to simplicity of \HI\ spectral lines).  As a result, about $41\%$ of the \bighicat\ sample probes environments with $A_V<0.2$. 
In Figure~\ref{f:hi-gas-fraction} we also show gas fraction histograms by splitting the sample into $A_V <1$ and $A_V >1$ (shown in blue and orange lines, respectively). The $A_V <1$ sub-sample peaks at 80-100 K, while the $A_V >1$ sub-sample peaks at a lower temperature ($\sim 50$ K).
This figure suggests that the broad CNM distribution comes from a combination of denser and more diffuse environments, with different physical environments resulting in the different range of temperatures.

Overall, we conclude that the observed CNM $T_s$ distribution is in line with theoretical expectations. It is broad, but the width appears in line with expectations for the environmental dependence of $T_s$, as well as how local turbulent perturbations affect \HI\ temperature. Building larger CNM samples to probe even more diverse ISM environments is highly important for the future. 

\subsubsection{CNM and H$_2$ temperature}
Because it is expected that molecular hydrogen (H$_2$) forms largely out of the CNM \citep[e.g.,][]{krumholz09} an important question is: how do the CNM and H$_2$ temperatures compare? For example, 
\cite{heiles03b} compared the CNM temperature with the H$_2$ temperature measured by FUSE for four directions where radio and UV sources were spatially close. Their conclusion was that H$_2$ and CNM temperatures did not agree. 
If we perform a similar analysis using the sightlines in the \bighicat{} within $\sim1^\circ$ of FUSE sightlines in \citet{shull21}, we find that the H$_2$ temperature, as measured through the lowest three rotational states (J = 0, 1, 2) of H$_2$, always lies between the minimum and maximum CNM temperature along the line of sight. In 3/4 cases, the H$_2$ temperature is consistent with the temperature of the CNM component with the highest \HI{} column density. 
Yet, as \citet{heiles03b} noted, the temperatures derived for H$_2$ and \HI{} are not directly comparable. The H$_2$ temperatures are calculated as a weighted mean over all velocity components because of the strong line saturation.

The mean excitation temperatures in \citet{shull21} are
$\langle T_{01} \rangle = 88 \pm 20$ K and $\langle T_{02} \rangle = 77 \pm 18$ K, which is very close to the CNM peak seen in Figure~\ref{f:hi-gas-fraction}. For sight lines
with $E(B-V ) > 0.5$ (corresponds to $A_V>1.5$) and $N_H > 10^{20.7}$ cm $^{-2}$, the H$_2$ temperatures decreased to 50–70 K.
This is qualitatively in agreement with Figure~\ref{f:hi-gas-fraction} for $A_V>1$ where we see that $T_s$ shifts to lower temperature, supporting the idea that H$_2$ and the CNM are thermally coupled. 
Detailed comparisons of larger samples of \HI\ velocity components with H$_2$ measurements 
remain as an important future task.
Homogeneous samples of other molecular species found in the diffuse ISM are starting to emerge thanks to ALMA and NOEMA. For example, \citet{rybarczyk22} found that the 
molecular gas traced by HCO$^+$, C$_2$H, HCN, and HNC is associated only with \HI\ structures that have an HI optical depth $> 0.1$, a spin temperature $< 80$ K, and a turbulent Mach number $>2$.
This result hints that not all CNM is useful for forming H$_2$, with only colder and denser CNM being 
thermally coupled to H$_2$.

\begin{figure}
	\centering
	\includegraphics[angle=-90,width=\textwidth]{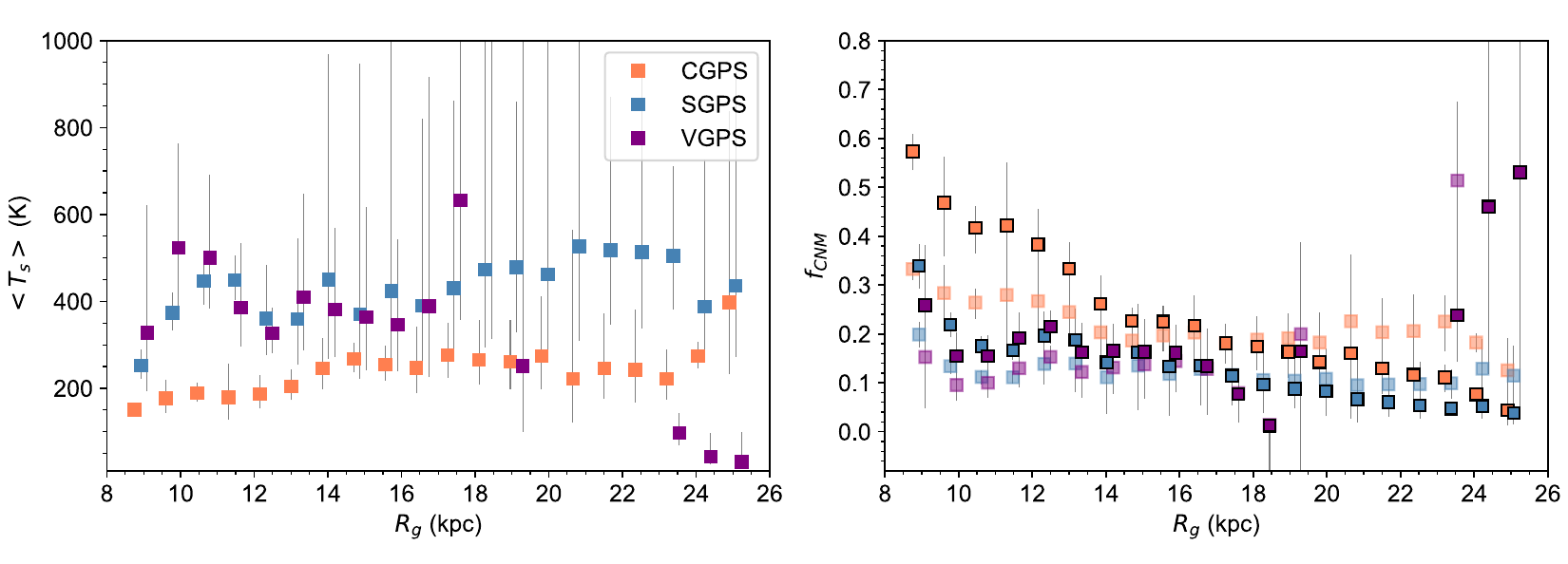}
	\caption{Left: line-of-sight mean spin temperature, $\langle T_s\rangle$, reproduced from data in \citet{dickey09}. The mean spin temperature is remarkably constant with Galactic radius in the outer Galaxy. Right: the implied CNM fraction $f_{\rm CNM} = T_s/\langle T_s \rangle$ as a function of galactic radius.   The CNM fraction shown here is dependent on the assumed cold gas temperature.  The dark points assume $T_s$ varies with radius based on \citet{wolfire03} (described in the text) and the lighter points assume $T_s = 50$ K.   The VGPS data are affected by the Galactic warp, which takes the cold disk out of the survey area.} 
	\label{fig:ts_vs_r}
\end{figure}

\subsection{CNM fraction}
As the CNM is the key building block for H$_2$, constraining how the CNM fraction varies across the Milky Way is essential for understanding star formation efficiencies. The Millennium Arecibo survey showed
that the majority of 79 observed directions had a CNM column density (or mass) fraction $<0.3$, with only a few lines of sight being dominated by the CNM \citep{heiles03b}.
In a  few directions, this study found essentially no CNM  and suggested that the CNM was destroyed in these directions by superbubbles.  
They also noticed that the CNM fraction increased with the total \HI\ column density up to $\sim 1.2 \times 10^{20}$ cm$^{-2}$, and then leveled off.  By comparing the WNM/CNM fraction at $|b|<10$ degrees, where lines of sight  are mainly in the Galactic disk, with higher latitudes, \citet{heiles03b} tested the hypothesis that the WNM fraction should be lower in the plane where thermal pressure is the highest but found no difference. 

With larger samples and more sensitive data, observers are starting to find regions with CNM fractions that clearly deviate from typical ISM values.  \citet{murray21} showed that  the Galactic \HI\ in the foreground of Complex C is particularly under-abundant in the CNM. 
The whole region has a relatively low column density ($<10^{20}$ cm$^{-2}$ on average) and in spite of narrow \HI\ linewidths shows a very low CNM fraction. 
Although the line-of-sight CNM fraction in the region is different from typical ISM fields, individual velocity components have similar 
$T_s$ and $\tau_0$,
suggesting that this region is particularly quiescent.  \citet{murray21} concluded that the region may be sampling an area that has not been recently disturbed by supernova shocks, leading its \HI\ properties to be dominated by thermal motions rather than nonthermal, turbulent motions.

At the other extreme, observations near several GMCs have shown on average higher CNM fractions. For example, 
\cite{stanimirovic14} investigated 26 \HI\ absorption and emission pairs
obtained in the direction of radio sources all in the vicinity of Perseus molecular cloud. 
Their CNM fraction is in the range 0.2-0.5, with the median CNM fraction being $0.33$ (in comparison to $0.22$ from \citealt{heiles03b}). Similarly, \cite{nguyen19} used 77 \HI\ emission-absorption pairs in the vicinity of Taurus, California, Rosette, Mon OB1, and NGC 2264 and found the CNM fraction in the range $\sim0.2-0.75$ with the median value of $0.35$.  These results suggest a scenario in which a high CNM fraction is required for molecule formation, and GMCs are built up stage by stage—from WNM-rich gas to CNM-rich gas to molecular clouds. Another reason for the high CNM fraction around GMCs could be effective CNM accretion. An important future task will be to provide even denser grids of \HI\ absorption sources and map spatially and kinematically the distribution of the CNM in the vicinity of GMCs.

The line-of-sight mean spin temperature $\langle T_s \rangle$ can also be used as an indicator of  the CNM fraction where $f_{\rm CNM} = T_s/\langle T_s \rangle$,  assuming $T_s$ as an average cloud temperature. 
\citet{dickey09} compiled \HI\ absorption-emission measurements in the Galactic plane from the three Galactic Plane Surveys (CGPS, SGPS and VGPS) to measure $\langle T_s \rangle$ as a function of Galactic radius, $R$.
We have reproduced those data in Figure~\ref{fig:ts_vs_r}.    We also show in Figure~\ref{fig:ts_vs_r} the implied CNM fraction $f_{\rm CNM} $ with two representative  assumptions of $T_s$:  average cold temperatures from \citet{wolfire03} heating and cooling models ($T_s= 85$ K at $R=8.5$ kpc to $T_s= 44.1$ K at $R=18$ kpc) or a constant temperature of $T_s = 50~{\rm K}$. If the typical cold cloud temperature decreases with Galactic radius then $f_{\rm CNM}$ also decreases from $\sim 0.3$ near the Solar circle to less than $0.05$ at $R\sim25$ kpc.  On the other hand, if we assume a flat $T_s$, as indicated by \citet{strasser07}, the CNM fraction is remarkably flat to large radii.   Clearly the key observational element required to understand how the CNM fraction varies will be the typical cold cloud temperature.   For completeness we note that although these curves were calculated under the assumption of smooth circular rotation, which breaks down for individual sight-lines, the trends largely hold for the extended azimuthal ($>70\arcdeg$) averages of the data shown here.  

\HI\ emission surveys have also been used, together with spectral decomposition, to determine the fraction of gas in the CNM. For example, \citet{kalberla18} found that for Galactic latitudes $|b| > 20\arcdeg$ the column density fraction of CNM was 25$\pm 5$\% at local velocities, whereas  \citet{marchal21b} estimated that in their high latitude field the average CNM mass fraction was $8\pm6$\%, showing large excursions (up to $\sim 30$\%) along filamentary structures.  Using a convolutional neutral network applied to the large-area survey GALFA-HI \citet{murray20} found that the CNM fraction varies from $<10$\% to $\sim 40$\%, depending on sky location.

In the full \bighicat\ compilation the mean CNM fraction (of $T_s < 250$ K) along the line-of-sight is $0.35$ and the median is $0.34$.  However, the \bighicat\ compilation is not uniform and contains numerous fields selected to be near the Galactic plane or molecular clouds.  Limiting the sample to $|b|>20\arcdeg$ gives a mean of $0.31$ and a median of $0.27$ and might be more representative of the general ISM.  The \bighicat\ CNM mass fraction is 40\% if we assume that 50\% of the total \HI\ mass is detected only in emission \citep{murray18}.

\subsection{Environmental Dependencies of Cold \HI\ Properties} 
Theoretical models and numerical simulations show that the thermal pressure at which cold and warm \HI\ co-exist depend on key physical properties such as the ambient interstellar radiation field, metallicity, dust properties (composition, charge, efficiency etc), and interstellar turbulence. As these physical quantities vary across galaxies, we expect to see variations of \HI\ gas properties. For example, photoelectric heating by dust grains may be enhanced in
particularly dust-rich environments.
Large-scale gradual metallicity variations are common across spiral galaxies \citep[e.g.][]{shaver83}. Even small-scale metallicity variations could be common as suggested by \citet{deCia21} who showed 
that many regions in the Solar vicinity have low metallicity, down to about 17 per cent Solar metallicity and possibly below. 
Similarly, as shown in Figure~\ref{f:shmuel-plot}, the CNM temperature should be affected by dust shielding in higher $A_V$ regions.
However, besides several sporadic studies (some listed in the previous sections), the observational evidence of the regional diversity of CNM properties across the Milky Way is still lacking.  This is largely due to the limited number of \HI\ absorption spectra, as well as large measurement uncertainties in deriving key physical properties ($T_s$, $f_{\rm CNM}$). With the systematic observational and data analyses approaches, and larger data samples in recent years, some interesting trends are starting to emerge.

\subsubsection{Does the CNM temperature vary across the Milky Way?}
Over the years, sporadic observations have shown occasional directions with $T_s\sim20-30$ K \citep{knapp72,meyer06,meyer12}. These temperatures are very low and in the realm of what is usually found for molecular gas.  
Explaining theoretically such low temperatures requires the absence of photoelectric heating \citep[e.g.,][]{spitzer,wolfire03}, and/or significant shielding \citep{glover10,gong17}. On the other hand, many studies have noticed relatively uniform $T_s$ for the CNM in the inner and outer Galaxy \citep{strasser07}, in and around the Perseus molecular cloud and several additional GMCs 
\citep{stanimirovic14,nguyen19}.
Using line-of-sight integrated properties,
\cite{denes18} hinted at the existence of a slightly warmer $T_s$ close to the Galactic center, as did \citep{bihr15} near the W43 “mini-starburst" star-forming region. 

Theories of  heating and cooling balance (e.g.\ Figure~\ref{f:shmuel-plot}) show that regions with high dust column density (high $A_V$) have a low CNM temperature, whereas in low $A_V$ regions the CNM temperature takes on a wide range of values \citep{bialy19,bialy20}.  We use all available \HI\ absorption spectra compiled in the \bighicat\ dataset to test whether $T_s$ depends on the optical extinction.  As shown in Figure~\ref{f:hi-gas-fraction}, the \HI\ components in the directions with $A_V<1$ occupy the higher-$T_s$ portion of the distribution relative to \HI\ components in the directions with $A_V>1$ which predominantly sample the lower-$T_s$ portion of the $T_s$ distribution.  This demonstrates that higher $A_V$ regions provide more intense shielding, which result in lower $T_s$ (see also Figure 10 in \citealt{kim22}). While this result is in line  with theoretical expectations the observed distributions of the cold \HI\ are broader than theoretical expectations. This could be due to the presence of turbulence 
which spreads out the temperature distribution. 

Clearly, these observations are just the beginning and future \HI\ surveys with larger samples of absorption lines are needed to confirm these trends. These results  demonstrates the power of large \HI\ absorption samples to, for the first time, depict differences of \HI\ temperature distributions caused by the underlying physical conditions.

\begin{figure}[h]
	\includegraphics[width=\textwidth]{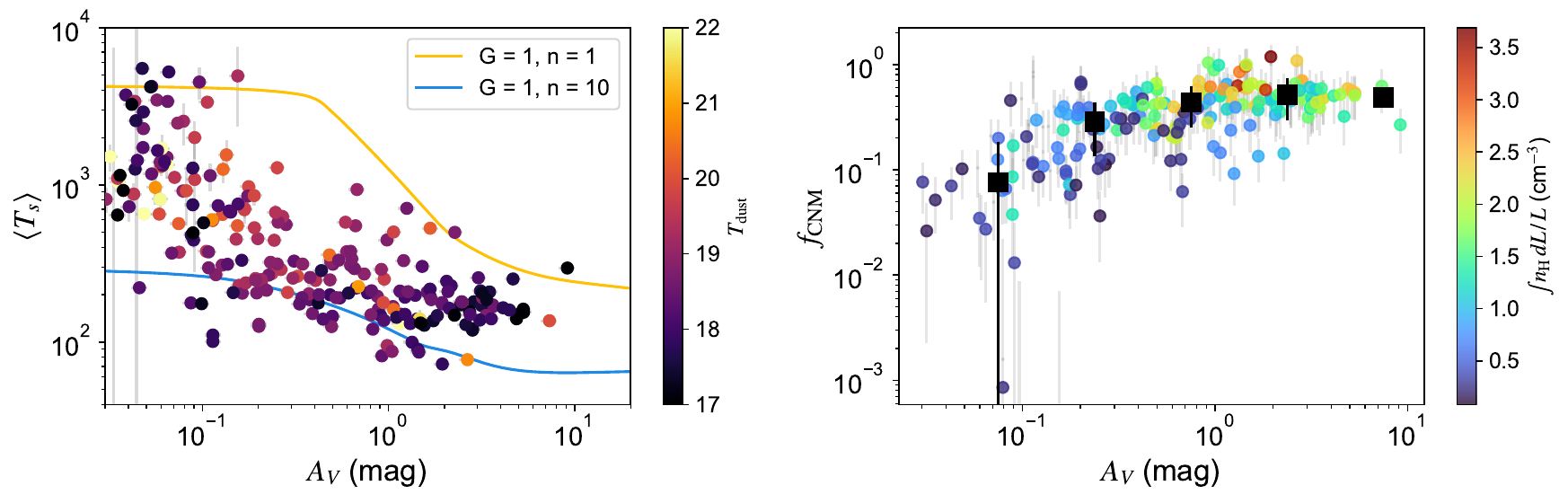}
	\caption{Left: The optical-depth weighted spin temperature from the \bighicat\ datasets as a function of optical extinction from Planck. Points are coloured by the dust temperature and lines are from the PDR model by Gong et al.\ (2017).  Right: The CNM fraction from the \bighicat\ dataset. The points are colored by mean volume density $\langle n_{\rm{H}} \rangle$, extracted from the 3D \citet{lallement19} dust map. The \HI\ absorption data probe arcsecond angular scales, while $A_V$ from Planck probes angular scales of $\sim5'$. The high-$A_V$ limit in this figure is not well defined as \HI\ spectra can be saturated and complex for interpretation.}
	\label{f:Ts_vs_Av}
\end{figure}

\subsubsection{What physical conditions are essential for setting \HI\ properties?}
The $T_s$ distribution in Figure~\ref{f:hi-gas-fraction} suggests that the \HI\ temperature depends on optical extinction.   To investigate a physical origin of the dependence of temperature on extinction in Figure~\ref{f:Ts_vs_Av} (left) we show the optical depth-weighted average spin temperature, $\langle T_s \rangle = \int \tau (v) T_s (v) dv/ \int \tau (v) dv $, as a function of $A_V$. Both $\langle T_s \rangle$ and $A_V$ are line-of-sight quantities.  We see that cold \HI\ exists over a very broad range of conditions: from very diffuse environments $A_V\sim0.03$ all the way to $A_V\sim10$, which is often used to characterize dense molecular clouds \citep{snow06}. 
However, the observed distribution is smooth, suggesting a continuous range of \HI\ structures, instead of well-segregated groups (e.g.\ diffuse atomic, diffuse molecular,  translucent, dense molecular). 
We overplot in this figure two color lines showing predictions from the PDR model from \cite{gong17} for particular input values $G\sim1$ (i.e.\ solar UV radiation field) and $n=1-10$ cm$^{-3}$. These lines nicely bracket the observed $\langle T_s \rangle$ values showing that the 
local density plays a crucial role in
constraining \HI\ properties. 

As discussed above, $\langle T_s \rangle$ is related to the CNM fraction along the line-of-sight, $f_{\rm CNM}$\footnote{The values of $f_{\rm CNM}$ shown in Figure \ref{f:Ts_vs_Av} are calculated using Equation 18 of \citet{kim14}. Uncertainties are calculated by varying the warm and cold cloud temperatures, as in \citet{murray21}. We find the same qualitative results if we take $f_{\rm{CNM}}$ as $N|_{T_S<250~\rm{K}}/N_{\rm{LOS}}$.}  as $\langle T_s \rangle \sim T_s/f_{\rm CNM}$ so the trend of $\langle T_s \rangle$ shown in Figure~\ref{f:Ts_vs_Av} (left) can reflect  variations in the cold cloud temperature, $T_s$, or the CNM fraction, $f_{\rm CNM}$.
However, for $A_V>0.2$ (see Figure~\ref{f:shmuel-plot}) the CNM temperature has a very narrow distribution ($\sim50$ K).  Assuming a well constrained CNM temperature for $A_V > 0.2$,  Figure~\ref{f:Ts_vs_Av} (right) shows that $f_{\rm CNM}$  increases with $A_V$ from $\sim10$ to $\sim80$\%.
The color of the points in the right panel of Figure \ref{f:Ts_vs_Av} shows the mean volume density along the line of sight,
showing that the sightlines with higher $A_V$---and higher $f_{\rm{CNM}}$---are those with the highest mean densities. We conclude that local density plays a crucial role in  constraining the CNM fraction. Higher local density implies more shielding, as well as cooling, and therefore more CNM.

While some particular conditions could result in the \HI\ being entirely  CNM, this is not the case in the Milky Way. The mean points in the right panel of Figure \ref{f:Ts_vs_Av} (shown as black squares), show that the CNM fraction flattens at about 60-80\%. 
Some of the highest-$A_V$ directions certainly probe environments around dense molecular clouds so the flattening of the CNM fraction at $A_V\sim1$ could signify the transition of \HI\ into H$_2$.  However, as \cite{stanimirovic14} pointed out, even lines of sight that probe deep inside the GMCs have a contribution from the WNM. The geometry and the level of mixing of the CNM and WNM are still not understood;
it is not clear if the WNM is located primarily in outer regions of the \HI\ envelope as theorized by \citet{mckee77} or if WNM is well-mixed by turbulence.  A priori one might assume that WNM mixed into high-pressure molecular clouds would rapidly cool, but \citet{hennebelle06} showed that the dissipation of
magnetic waves can provide substantial heating to  maintain the
WNM inside molecular clouds. 
It therefore seems unsurprising that the CNM and WNM are well mixed in all environments and $f_{\rm CNM}<1$.     

The low-$A_V$ end of Figure~\ref{f:Ts_vs_Av} is also interesting, but poorly constrained by the observational data. \citet{kanekar11} noticed that sightlines with column density $>2 \times 10^{20}$ cm$^{-2}$ had low optical-depth averaged spin temperature, which they interpreted as a evidence of a threshold for CNM formation. \cite{kim14} suggested that instead of a column density threshold, $A_V\sim0.1$ may simply correspond to a characteristic length scale for CNM structures.  In spite of the larger sample in \bighicat, the jury is still out on whether column density threshold exists for CNM formation.

\subsection{The UNM temperature and fraction}
It has been difficult to constrain the observational properties of  \HI\ at temperatures $T_s>250~{\rm K}$.  Until recently most estimates of the UNM temperature were made as upper limits from the line-width-based kinetic temperatures of either emission or absorption.  UNM mass fraction estimates made from \HI\ emission give fractions between $28\pm11\%$ \citep{marchal21b} and $41$\% \citep{kalberla18}.  Using upper limits to kinetic temperatures of \HI\ absorption, \cite{kanekar03}  found UNM fractions of: 77\% and 72\% by mass for two LOS.  By contrast, \citet{begum10}  estimated $27\%$ across five lines-of-sight and \citet{roy13b} set a lower limit of $\sim 28$\% mass fraction with values as high as $44\%$. The lower limit is based on the conservative assumptions that all detected CNM has $T_s=200\rm\,K$ and all non-detected WNM has $T_s=5000\rm\,K$, and not on direct measurements of the absorbing and emitting properties of the unstable gas.  Limits on the fraction of the medium in the unstable state based on spin temperature should be more robust, but have been difficult because of the low optical depths inherent to the UNM and WNM.

Only a handful of observations have achieved the optical depth sensitivity required to measure $T_s$ in the UNM, e.g., \citet{heiles03b} detected 13 UNM components giving an estimated column density fraction of 29\% for spin temperatures measured out of the plane.  The best estimates for UNM fraction have come from the 21-SPONGE survey \citep{murray18}, which conducted very deep \HI\ absorption observations 
and was sensitive to \HI\ with spin temperature up to $\sim4000$ K.
Not only this was the first statistically significant survey designed to detect low optical depth ($\tau \sim \mathrm{few} \times 10^{-3}$) broad lines, but this survey also took exceptional care of observational biases introduced in the analysis method used to constrain \HI\ spin temperature and recover the \HI\ mass distribution.  21-SPONGE  tripled the number of UNM components and found $20\%$ of the total \HI\ mass, and $41\pm10$\% of the \HI\ mass detected in absorption, to correspond to the UNM \citep{murray18}.  The result agreed  with the \citet{heiles03b} 
sample and the \citet{begum10} and \citet{roy13b} studies.  The high sensitivity \citet{nguyen19} survey subsequently found a similar fraction.  


A seemingly trivial, but important, distinction in the various observational and theoretical studies is the chosen boundary between the CNM and UNM. By combining several \HI\ absorption surveys in \bighicat, we have the largest database of high latitude ($b>15\arcdeg$) spin temperatures to search for a possible observational boundary between CNM and UNM.  We find no clear observational boundary,
but suggest that observers and theorists adopt the value of $250$ K as an nominal boundary between CNM and UNM simply to ease cross-comparison.

Driven largely by 21-SPONGE, the consensus seems to be pointing towards about 20 - 30\% of the \HI\ by mass being in the unstable phase.  The result is similar to the \HI\ emission derived UNM fractions (28\% and 41\%) by \citet{marchal21b} and \citet{kalberla18} given the errors on those estimates. The observed UNM fraction agrees well with the mass fraction at UNM temperatures from the TIGRESS and SILCC numerical simulations shown in Figure~\ref{fig:Ts_TIGRESS} \citep{kim17,rathjen21}. 

The spatial distribution of the UNM is currently uncertain.
For example, \cite{murray18} noticed that  highest Galactic latitudes are dominated by WNM, while the CNM and UNM dominate 
low Galactic latitudes. However, the Gaussian decomposition is least reliable at low latitudes.  On the other hand, \cite{kalberla18} who studied UNM via \HI\ emission lines, concluded that CNM structures are surrounded by the UNM which often has filamentary morphology. As shown in Figure 5, from the \bighicat\ catalog we see that regions with $A_V>1$
have slightly colder peak $T_s$, due to increased shielding, and also exhibit less UNM (at $T_s \sim 200-600$ K) relative to the more diffuse \HI\ environments with $A_V<1$. This is likely caused by more pronounced dynamical processes at low-$A_V$, and more shielding at high-$A_V$, in agreement with suggestions by \cite{wolfire15}.
Future observational constraints of the UNM spatial distribution will be very important to  help determine the processes that drive \HI\ out of equilibrium.

\subsection{The WNM temperature and fraction}
As the WNM usually dominates \HI\ emission spectra, its properties are sometimes  derived purely from \HI\ emission data.  \cite{kalberla15,kalberla18} decomposed the HI4PI survey spectra into Gaussian components and selected three phases (CNM/UNM/WNM) based on the width of Gaussian functions. They found that the WNM was smoothly distributed with a mass fraction of $\sim 32$\%.
Similarly, \citet{marchal21b} decomposed the high resolution $12 \times 12$ degree  GHIGLS field into Gaussian components and found that  64\% of the \HI\ mass is in the WNM.  
However,  the standard deviation of that value is relatively high (35\% of the median value).  By comparing \HI\ with extinction data, they estimated the  WNM temperature $T_k = (6.0 \pm 1.3) \times 10^3$ K.

To robustly estimate the excitation temperature of the WNM, without a corruption by turbulent motions, absorption-emission pairs are required.
High sensitivity interferometric \HI\ absorption observations have  been used to measure the WNM temperature. For example, \citet{carilli98}  measured $T_s = 5500-8700$ K in the direction of Cygnus A. \citet{dwarakanath02} and \citet{kanekar03}  found lower values of 3000-3600 K. 

The very sensitive \HI\ absorption survey, 21-SPONGE, detected a handful of components with $T_s > 1000$ K \citep[e.g., $<10$\% by number in][]{murray15}. To improve sensitivity to shallow, broad absorption features further, \cite{murray14} employed a spectral stacking
analysis on 1/3 of the survey data and detected a pervasive population of WNM gas with $T_s =7200_{-1200}^{+1800}$ K. This result was confirmed using data from the entire 21-SPONGE survey in \citet{murray18},  where residual absorption was stacked and binned based on residual emission, demonstrating the existence of a significant absorption feature with a harmonic mean $T_s\sim 10^4$ K. 

The range of predicted kinetic temperatures for the WNM 
from the most detailed ISM heating and cooling considerations,
is $T_k \sim 5000–8800$ K \citep{wolfire03}. If the \HI\ is only collisionally excited, as shown by \cite{kim14}, this range implies  $ T_s \sim 1000–4000$ K (their Figure 2). When, however, the common prescription is used to account for the Ly$\alpha$ flux, it is expected that $T_s\sim 3500–5000$ K.
The \cite{murray14} and \citet{carilli98}  results are 
significantly higher than predictions from standard ISM models based on collisional \HI\ excitation. 
As can be seen in Figure~\ref{f:hi-gas-fraction}, even all combined \HI\ absorption measurements together are not enough to detect the WNM peak of the \HI\ gas distribution.
This suggests that the WNM spin temperature is likely higher than standard analytical and numerical predictions.
Another problem could be the implementation of the $T_k$ to $T_s$ conversion and the use of a constant   Ly$\alpha$ flux.
In a recent study by 
\citet{seon20}, where a multi-phase model was used 
and photons originating from H{\sc ii} regions were tracked to produce the Ly$\alpha$ radiation field, the Ly$\alpha$ flux was strong enough to bring the 21 cm spin temperature of the WNM close to the kinetic temperature. This encouraging result demonstrates that a careful treatment of Ly$\alpha$ photons needs to be incorporated in numerical simulations, instead of commonly used uniform Ly$\alpha$ flux. 

In summary, the WNM remains difficult to detect in absorption and requires long integration times or stacking analyses. The detection difficulty likely stems from the possibility that it has a higher excitation temperature than what was expected.  Future surveys of \HI\ absorption will be essential to probe spatial variations of the WNM temperature via stacking.  Constraining methods that use Gaussian decomposition to separate \HI\ phases using simulated data is essential as studies of external galaxies rely on this method exclusively.

\subsection{The ``Dark" Neutral Medium }
The transition from \HI\ to  \htwo\ is a critical step in the evolution of galaxies.  The efficiency with which this transition occurs appears to influence star formation efficiency globally.  While the transition from \HI\ to \htwo\ is extremely important, it is also not easy to study.  \htwo\ is impossible to observe directly in typical star-forming conditions \citep{carilli13}.  Carbon monoxide (CO) is traditionally used as a proxy for \htwo, but because it cannot self-shield it is easily dissociated.  In the dense conditions of star-forming molecular clouds \htwo\ and dust shield CO so that it is readily detectable, but in the early stages of the atomic-to-molecular transition CO is not observed. 
At the same time, the CNM that goes on to make H$_2$ is studied via absorption measurements that are largely limited to sight-lines in the direction of background sources. 
Excluding the dominant tracers of CO and \HI, how do we find gas in its transition from \HI\ to \htwo?

The Fermi $\gamma$-ray telescope and the Planck telescope independently discovered a component of so-called ``dark'' gas, not detectable by \HI\ or CO but inferred by its excess $\gamma$-ray emission and dust opacity \citep{grenier05, planck11}. Astonishingly, the ``dark'' gas mass seems to be  between 20 and 50\% of the total \HI\ and CO detected gas in the solar neighborhood.   It was thought that this might be the vast reservoir of transition gas, caught between its purely atomic and self-shielding molecular states.  The discovery prompted a flurry of research into whether the ``dark" gas was primarily optically thick atomic gas or primarily CO-dark molecular gas \citep[e.g.][]{fukui14,fukui15,lee15,murray18,sofue18,nguyen19}.  

To resolve the nature of dark gas, indirect tracers are often used.  For example, \citet{fukui14,fukui15}  assumed that, if dust and neutral gas are well mixed and that the specific dust opacity is constant, the \HI\ optical depth and spin temperature can be approximated from the observed Planck  dust properties. Under these assumptions they estimated that the \HI\ optical depth must exceed one along most local sight-lines but that low resolution data missed significant unresolved dense gas.  But conversely, \citet{lee15} used direct measurements of \HI\ absorption to show that in the Perseus molecular cloud less than 20\% of sight-lines show $\tau > 1$.   \citet{murray18} revisited this problem using some of the techniques of \cite{fukui15} applied to higher angular resolution GALFA-HI data and found that the \HI\ is largely diffuse at high latitudes down to 4\arcmin\ scales and the lower angular resolution observations are not missing optically thick \HI. \citet{nguyen18} showed that the \HI\ opacity effects only become important above $N_{\rm HI} > 5 \times 10^{20}~{\rm cm^{-2}}$.  

Meanwhile, several studies have investigated alternative molecular gas tracers (HCO$^+$ \& OH) instead of CO.  \citet{liszt18} observed HCO$^+$---known to trace H$_2$ even in diffuse environments---in addition to CO towards 13 background sources in the direction of the Chamaeleon cloud. Whereas CO emission was detected in only 1 of these directions, HCO$^+$ absorption was detected in 12/13 directions, and the H$_2$ column densities inferred from HCO$^+$ explained the discrepancy between the column densities derived from $\gamma$ rays and dust emission with those estimated from \HI{} and CO. While they found that optically thick \HI{} could contribute to the dark gas in several directions, their results showed that the dark gas could primarily be accounted for by dark molecular gas rather than dark atomic gas (although \citealt{hayashi19} presented a contradictory view that Chamaeleon has more optically think \HI\ rather than CO-dark \htwo).  Independently, \citet{nguyen18} showed that the OH could be used as a proxy for \htwo\ over a broad range of column densities and that optically thick \HI\ is not a major contributor. In the end, it seems that a combination of factors including a small amount of optically thick \HI\ ($\sim 10-30\%$ increase in column density), evolution of dust properties and CO-dark molecular gas can resolve the dust opacity excess.

\section{CHARACTERIZING THE STRUCTURE OF  H{\sc i} ON ALL SCALES}
\label{sec:structure}
The morphology of the \HI\ in the Milky Way is primarily studied through \HI\ in emission and therefore our knowledge of structure has been dominated by the WNM. Observational advances with spatial and spectral resolution over the past twenty years have started to reveal some hints about the CNM structure.  Extending  our understanding of the statistical thermal properties of the two stable phases to an understanding of the differences in morphology of the Galaxy's structure in the \HI\ phases is still underway.    

The structure of \HI\ in emission can be characterized in two ways: {\em deterministic} and {\em scale-free}.  Through the ever-improving observational data, supported by numerical simulations \citep[e.g.][]{kim18,hill18}, a picture has developed of the \HI\ gas cycle in which the diffuse, scale-free \HI\ is repeatedly ordered by large-scale compression into dense, often filamentary and magnetized \citep{mcgriff06b}, deterministic structures that catalyze cooling \citep{dawson11a} and may drive turbulence.   Deterministic structures are those that likely  originate from a specific event or series of events; of those the most common are the shells, bubbles and chimneys that  dominate the \HI\ sky on angular scales between one and hundreds of degrees \citep{heiles79,hu81,heiles84,mcgriff02a,ehlerova05}.

\subsection{Deterministic Structure: From Bubbles \& Shells to Clouds, Filaments, \& Chimneys}
While some of the \HI\ shells observed in the Milky Way are clearly associated with relics of massive stars \citep[e.g][]{normandeau96,mcgriff01c}, many of the largest  \HI\ shells have not been clearly connected to their parent stars.  In fact,  
\HI\ shells are rarely observed around currently active star-forming regions or even young supernova remnants.  It seems that the \HI\  shells are more often associated with larger, older objects, although there are some notable exceptions,  e.g.\ GLIMPSE bubble N107 \citep{sidorin14}.   

Identifying \HI\ shells in the Milky Way is not easy.  Until recently the majority of shells  had been cataloged manually \citep{heiles79,heiles84,mcgriff02a}. Some attempts had been made to catalog shells semi-autonomously \citep[e.g.][]{ehlerova05} with varying success.  Issues affecting the success of automated searches relate to the velocity ambiguity of the inner Galaxy, the prevalence of arcs or shell-like features even in turbulent data (a quick look at terrestrial clouds can confirm!) and complex overlapping structures.  Most recently, \citet{ehlerova13} refined automated shell searches with requirements for completeness of the shell, velocity widths, etc.  The resultant catalog from the LAB survey has 333 entries, distributed across the sky.  Considering only shells outside the solar circle, the fraction of the disk's surface area covered by shells drops from $\sim 80$\% at the solar circle to $\sim 10$\% at 15 kpc.  Apart from a handful of newly identified shells, work on \HI\ shells has largely stagnated over the last decade.  Progress has been partially limited by the lack of distances to \HI\ structure and the difficulty in defining three-dimensional structures.  This is set to change in the coming years with three-dimensional dust maps helping to overcome problems with velocity ambiguities. Despite the difficulties in robustly identifying \HI\ shells, they do have important roles in the evolution of \HI\ structure and density.  

From the various observational studies published on \HI\ supershells throughout the late 1990's and 2000's it became clear that the role of supershells is to order the diffuse \HI\ into large-scale, filamentary structures visible throughout the Galactic plane.  Does the \HI\ also cool within these structures?  Simulations  have shown that  supershells sweep up the warm, ambient medium into dense, structured filamentary walls where the gas can cool to form CNM and molecular clumps \citep{koyama00,maclow05,hopkins12}.  From an observational perspective this role is less well documented although there certainly have been examples of supershells with associated molecular gas \citep[e.g.][]{fukui99,dawson08}. \citet{mcgriff03} found  clumps of narrow line-width \HI\ in the walls of a large-supershell, suggesting that the structures were formed by Rayleigh-Taylor instabilities.  \citet{knee01} and \citet{moss13} each used HISA to show that the walls of two  supershell structures were dominated by cold gas. The most comprehensive analysis of supershells as collectors and coolers of gas was produced by \citet{dawson11a}, who showed that within two supershells the walls were dominated by cold gas with estimated temperatures and densities of $T_k\sim 100~{\rm K}$ and $n_H\sim10~{\rm cm^{-3}}$.  Comparing the walls to the surrounding medium, they found statistically significant indicators of narrow linewidths consistent with enhanced CNM fraction as well as molecular fraction.

{\bf Filamentary structures.} Feedback events, like those observed in supershells, indicate that star-formation helps establish the large-scale structure of the \HI\ that may ultimately be devolved to smaller scales. Much of that large-scale structure after the star formation tracers are gone is filamentary. For example, \citet{soler20} used the THOR survey at  40\arcsec\ angular resolution to show that within the Galactic plane the \HI\ has a filamentary nature with the filaments predominantly aligned with the Galactic plane.  In some places where star formation appears to have been active, the filamentary structure abruptly switches direction to out of the plane.  Extending this work across the Galactic disk, \citet{soler22} showed that in the outer Galaxy the filamentary \HI\ structures are predominantly parallel to the Galactic plane and inside the Solar circle they are mostly perpendicular to the plane or without a preferred orientation. Their interpretation is that the filaments show the imprint of supernova feedback in the inner Galaxy and Galactic rotation and shear in the outer Milky Way. 

Observationally, distinguishing the characteristic morphology of the CNM from that of the WNM has been difficult to disentangle because the large linewidth and diffuse character of the WNM blurs structure within position-position-velocity cubes of \HI\ emission.  The morphology of the CNM, with its generally smaller brightness temperatures  and narrow linewidths, is easily obscured within spectral channels dominated by warm gas.   But spectacular examples of discrete, cold features that appear cloud-like or filamentary abound.  For example, the very local Leo cold cloud's  exceptionally narrow velocity width indicates a kinetic temperature of  between 11 and 20 K (Peek et al.\ 2011;  supported by absorption measurements (Murray et al.\ 2018)).\nocite{peek11b}  
Recent work making use of the combination of spectral and angular resolution, and sensitivity that have been afforded by the GALFA-HI survey has revealed small-scale, narrow linewidth, fibrous structure \citep{clark19a}.  The  inference that the fibrous structures are cold has been confirmed through various techniques including correlation with NaI absorption \citep{peek19}, convolutional neural network CNM separation \citep{murray20} and dust-to-gas comparison \citep{kalberla20}.  All three papers hint that the morphology of the CNM, as observed in emission, is distinct from the WNM.  Dense grids of \HI\ absorption spectra, as well as new techniques that perform spectral decomposition of \HI\ emission for separating the WNM and CNM by using \HI\ absorption data as learning templates \citep[e.g.][]{murray20}, 
will be important for further characterizing the CNM morphology.

{\bf Disk-halo clouds.} The familiar  paradigm that supershells can drive hot gas and metals is clearly illustrated by many examples of \HI\ supershells that completely disrupt the \HI\ disk \citep{normandeau96,heiles98,mcgriff02a,pidopryhora07,dawson08}.   They also seem to contribute to defining of the structure the disk-halo interface region into ``clouds" or ``cloudlets".  In their 1990 review of \HI\ in the Milky Way \citet{dickey90} asserted that the concept of \HI\ clouds was not a useful one.  Indeed, a quick examination of \HI\ emission images, such as those shown in the movie in the Supplementary material will confirm that assertion - how could any of this structure be defined as a ``cloud"?  
Nevertheless, this description of \HI\ has persisted in the ISM lexicon.   In spite of its general appearance there are a few examples of warm \HI\ in the disk-halo interface region that can reasonably be described as ``cloud"-like.  \citet{lockman02a} identified an unexpected population of  \HI\ structures that are clearly discrete and bounded and therefore cloud-like. These parsec-scale clouds are near the tangent point of the inner Galaxy, where small offsets from circular velocity place them at nearly forbidden velocities.  
Subsequent studies suggest that the disk-halo cloud population is not completely ubiquitous throughout the Milky Way \citep{stil06,stanimirovic06,ford08,ford10,pidopryhora15} but instead they may be a result of ISM structure driven in some indirect way from star formation.  \citet{ford10} connected the clouds with broken remnants of shells and supershells driven away from the disk.

\begin{figure}
	\includegraphics[width=\textwidth]{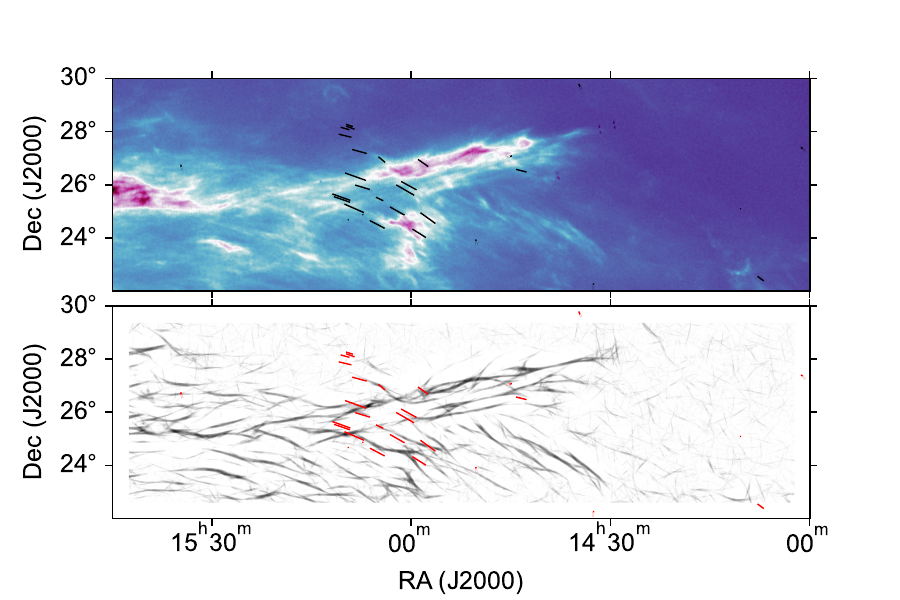}
	\caption{Top: \HI\ image at $v=-7~{\rm km~s^{-1}}$ from GALFA-HI \citep{peek18} showing fibrous structure overlaid with starlight polarization vectors \citep{heiles00}.  Bottom: Rolling Hough Transform  \citep{clark14} backlit image  of filamentary \HI\ structure from the same \HI\ data, overlaid with the same polarization vectors. 
		\label{fig:HI+magnetism}}
\end{figure}  

\subsection{Linking magnetic fields to \HI\ structure}
\label{subsubsec:magnetism}
Although  the link between atomic gas and magnetism is not necessarily intuitive, early work on the correlation of starlight polarization with many large-scale local  \HI\ structures,  such as Loop I and the Orion-Eridanus bubble \citep{heiles97}, indicated that magnetic fields are associated with the morphological structure of \HI.  Using observations of Zeeman splitting of the  \HI\ line in absorption, \citet{heiles89} showed that for a sample of lines-of-sight through local supershells the magnetic field strengths were larger than the typical ISM values.    The implication from the Zeeman work and the correlation of starlight polarization angle with \HI\ loops is that the magnetic field and \HI\ were co-structured through a common formation mechanism, such as a supershells \citep{heiles89,heiles98}.  

\HI\ Zeeman work culminated with the Millennium survey of \HI\ absorption, giving 69  measurements on arbitrary lines-of-sight with magnetic field strengths of $|B_{los}| \leq 10~{\rm \mu G}$ \citep{heiles04,heiles05}.  Through comparison of $|B_{los}|$ with column density they showed that the field strengths were not consistent with flux freezing, as is typically found in molecular clouds \citep{crutcher99}.   The technical challenges of \HI\ Zeeman observations have limited its extensive application and large-scale conclusions about the connection of magnetic field strengths with the structure of the neutralmedium. 

Over the last decade there has also been a growing body of work indicating that \HI\ structure  is linked to magnetic field structure, as traced by starlight and dust polarization \citep[e.g.][]{mcgriff06b,clark14,clark19b}. Figure~\ref{fig:HI+magnetism} shows an example \HI\ velocity channel image with the fibrous structure emphasized by the Rolling-Hough Transform overlaid withstarlight polarization.  
So far, most of the observational evidence linking \HI\ structure to magnetic fields has focused on the CNM.  Clearly the \HI\ Zeeman measurements, by nature of tracing absorption, exclusively measure CNM.   This connection with the CNM was borne out in \citet{clark19a} and \citet{kalberla20}, who showed that the magnetically correlated fibrous structure of the \HI\ is dominated by narrow linewidths.   Similarly, the structure of the cold Riegel-Crutcher HISA cloud (discussed in Section 4.3) was shown to have outstanding agreement with  starlight polarization \citep{mcgriff06a}.  Extending to the whole sky, \citet{kalberla21} found remarkably good agreement between far-infrared polarization angle and \HI\ filament orientation angle for velocity widths of $1~{\rm km~s^{-1}}$, but that for larger velocity widths the agreement disappeared.  The  conclusion seems to be that \HI\ filamentary structure is cold and magnetically dominated with the field orientation parallel to the filaments.  

Turning the technique around, the fibrous  \HI\ structure has been used as a  proxy for direct observations of magnetic fields, adding a velocity dimension  to  the  observational technique of dust polarization.  \cite{clark19b} performed a full-sky decomposition of the fibrous \HI\ that led to a velocity resolved projection of the plane-of-sky magnetic field.   An alternative method proposed by  \cite{gonzalez-casanova17} and \citet{yuen17} utilizes the connection of turbulent motion to the magnetic field by using gradients of the velocity amplitude, which are expected to be perpendicular to the magnetic field direction, to trace the magnetic field orientation.  In a series of papers, \citet{hu21}  employed another technique focusing on the structure function of simulated \HI\ to reveal both the magnetic field direction and its strength.  The underlying correlation of \HI\ structure and magnetic fields is an area of great activity and will likely develop rapidly over the coming years. 

The fact that the two ISM elements - magnetic fields and CNM - correlate so well is suggestive that the magnetic field is either dynamically important in setting the  structure of  \HI\ in many environments or that the same dynamical events (e.g.\ supershells) that help  produce CNM also enhance magnetic field strengths.  While most observations linking magnetic fields and \HI\ have focused on starlight or dust polarization as the tracer of magnetic field, there have been several studies in the past few years that have used Faraday rotation of radio continuum as the magnetic field tracer \citep{van-eck19,bracco20b,turic21}.  Radio continuum studies may help to break the bias of magnetic fields traced in cold gas by dust, instead showing magnetic fields in the warm, ionized medium probed by Faraday rotation.  However, recent low radio frequency polarimetric observations of Faraday rotation 
have generally found morphological association with cold \HI\ structures.  For the first time, \citet{campbell22} used mid-frequency radio polarization Faraday rotation to show that the \HI\ appears clearly connected to the warm ionized medium via a common magnetic field. While they consider whether a compressive event  triggered the connection of the \HI\ and magnetic field, they cannot definitively conclude that it did. Future studies of Faraday rotation and \HI\ will be important to reveal whether magnetism is as crucial to the structure of the WNM as the CNM.

\subsection{Statistical description of the \HI\ distribution}
Statistical studies provide an important and complementary way to describe and characterize complex, scale-free structure seen in high-resolution \HI\ images.
As interstellar turbulence is known to be ubiquitous throughout the ISM and greatly responsible for the \HI\ distribution in galaxies, e.g. \citet{elmegreen04}, statistical descriptors of \HI\ are important for constraining the drivers and the nature of interstellar turbulence. 
In addition, the \HI\ turbulence constraints are essential for understanding the process of molecular cloud formation and evolution as studies of molecular clouds \citep[e.g.,][]{hennebelle12} show that most of the turbulent energy is injected from outside.

Turbulence can be driven by many energy sources that operate on very different spatial scales. The key candidate is stellar feedback --  a term used to express the combined effect of proto-stellar jets, 
stellar winds, photo-heating, and supernovae \citep{offner14,grisdale17}.
Turbulent drivers operating mainly on larger spacial scales include
gravitational instabilities \citep{elmegreen03,krumholz16}, thermal instabilities \citep{kritsuk07,kimkim13,iwasaki14}, and magneto-rotational instabilities \citep{piontek04,piontek05,piontek07}.  
However, clear observational signatures of different drivers of turbulence are still lacking, partially due to 
the need for extremely high spatial dynamic range (from few pc all the way to many kpc). 

Many different statistical approaches have been used on \HI\ observations to characterize complex structure. Some of these approaches were summarized in \citet{elmegreen04} and \citet{scalo04}.  
Here we focus on the spatial power spectrum of \HI\ as one of the common tools used to analyze \HI\ emission, with a goal of compiling published results and pulling together some common trends.

\subsubsection{\HI\ spatial power spectrum (SPS)}
\label{subsubsec:SPS}

\begin{figure*}[h]
\vspace{-0.5cm}
	\includegraphics[scale=0.65]{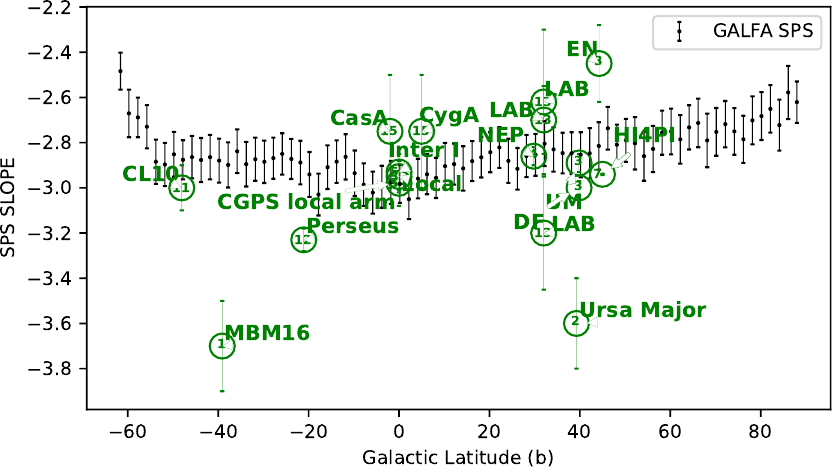}
	\\
	\caption{The SPS slope obtained for different \HI\ regions. All data points were derived from the \HI\ column density images integrated over a certain range of velocities, mostly focusing on the low-velocity HI. Black data points are from Mittal et al. (in prep). Green: \citet{pingel13}, \citet{pingel18}, \citet{choudhuri19}, \citet{marchal21a}, \citet{mivilledeschenes03}, \citet{blagrave17}, \citet{martin15}, \citet{khalil06}, \citet{dickey01}, \citet{kalberla19}, \citet{marchal21b}, \citet{green93}, \citet{crovisier83}, \citet{chepurnov10}.}
	\vspace{0cm}
	\label{f:SPS-slope-comparison}
\end{figure*}

In the presence of turbulence
the 2D spatial power spectrum (SPS) of \HI\ intensity or column density fluctuations is expected to have a single or double power-law functional form \citep{scalo87}.  The SPS of a 2D image, $I(\mathbf{x})$, is defined as 
\begin{equation}
P(k) = \int \int \langle I(\mathbf{x}) I(\mathbf{x}^\prime)\rangle \exp^{i\mathbf{L}\cdot k} d\mathbf{L},
\end{equation}
where $k$ is the spatial frequency and $\mathbf{L} = \mathbf{x}-\mathbf{x}^\prime$ is the distance between two points in the image.
The power-law slope of the SPS, $P(k) \propto k^{\gamma}$,  derived from 2D \HI\ images reflects contributions from  
the underlying 3D velocity and/or density fluctuations \citep{lazarian00}. 
Theoretically, the SPS can illuminate the spatial scales at which energy 
is injected or dissipated, as well as the mechanisms by which that energy 
cascades from large to small scales or vice versa. 

Practically, however, the interpretation of SPS results is complex due to many observational and theoretical effects. 
The SPS slope depends on
the line-of-sight complexity (e.g.\ multiple \HI\ structures in the same direction), the multi-phase nature of \HI, and optical-depth effects. Furthermore, the disk geometry and the \HI\ scale height can result in the transition between 2D and 3D turbulence, producing breaks in the SPS. 
Recent studies have shown that  the interpretation of SPS breaks can be greatly affected by telescope instrumental effects (beam and noise), and the multi-phase nature of the medium \citep{koch19,kortgen21}. Similarly, 
it has been shown that  separating density and  velocity fluctuations from observed intensity fluctuations requires a careful consideration. The approach for doing this by varying the width of velocity channels  \citep{lazarian00} has been recently updated with a more complex treatment by \cite{yuen21}. However, some debate 
still exists. \citet{clark19a} and  \citet{kalberla16} suggested that \HI\ intensity fluctuations seen at different velocities are phase dependent, with narrow velocity channels being dominated by the CNM and causing the \HI\ SPS slope to be more shallow than what is found for the integrated intensity images. 
Finally, it is currently not clear whether the SPS of \HI\ emission is sensitive in distinguishing between different turbulent modes within a multi-phase medium.

From the observational point of view, although there are many measurements of the \HI\ SPS, most studies employed different analysis techniques, resulting in heterogeneous samples 
that are hard to compare, making it difficult to search for spatial variations of turbulent properties with interstellar environments.
Early \HI\ SPS studies by \citet{crovisier83} and \citet{green93} fit a single power-law with indices of $-3.0$ and $-2.8$, respectively.  Subsequent large-area (many degrees across) \HI\ maps have also been used to calculate a single SPS. 
As many studies of \HI\ SPS  showed no evidence for a turnover on large spatial scales, the interpretation has been that \HI\ turbulence in the Milky Way is being driven on scales larger than the size of individual \HI\ structures, \citep[e.g.][]{pingel13,pingel18}. This is somewhat puzzling considering the existence of many star-forming regions and supernovae in the disk, which we expect would modify turbulent properties of \HI.

In Figure~\ref{f:SPS-slope-comparison} we attempt to compare values of the \HI\ SPS slope obtained from the literature. We note that all these measurements were obtained from \HI\ images integrated over a range of velocity channels that roughly corresponds to the local \HI\ ($\pm20-40$ km s$^{-1}$) and likely represent estimates of density fluctuations \citep[based on][]{lazarian00,yuen21}. We also  show  the application of the rolling SPS analysis on the entire region observed by the GALFA-HI survey from Mittal et al.\ (in prep). The rolling SPS method, introduced by \citet{szotkowski19}, calculates SPS locally by rolling a spatial kernel across an image. This allows for a consistent handling of instrumental effects.

The GALFA-HI survey covers about 13,000 square degrees and includes a region in the inner Galaxy, a region in the outer Galaxy, as well as some high-latitude \HI. The SPS slope, calculated over the angular range of scales from 1.5 degrees to 16 arcmin, varies between $-2.6$ and $-3.5$. Close to the Galactic plane, $|b|<10$ degrees, the SPS slope is the steepest, while at higher Galactic latitudes the SPS slope gradually becomes more shallow reaching to $\sim-2.7$ at $|b|>70$ degrees (Figure~\ref{f:SPS-slope-comparison}).
While the gradual change of SPS slope with Galactic latitude
cannot be explained with the optical-depth effects, 
it becomes less
significant when the \HI\ column density is corrected for the difference in the line-of-sight length caused
by the plane parallel geometry of the HI
disk. 


It is interesting to see that the \HI\ SPS slope appears relatively uniform across large spatial areas, and is only consistently slightly more shallow at high Galactic latitudes.
The GALFA-HI survey samples the range of \HI\ column densities from $\sim10^{20}$ to  $\sim10^{22}$ cm$^{-2}$, yet the SPS slope remains largely around $-2.8$ for $|b|<50$ degrees. 
This could be due to the dominance 
of large-scale turbulent driving, on Galaxy- or kpc-wide spatial scales, possibly caused  by a combination of Galactic rotation, gravitational instabilities, infall onto the Galaxy, and even energy injection from large supershells \citep{wada02,bournaud10,krumholz16}. Numerical simulations by \citet{yoo14} showed that the turbulent energy spectrum is very sensitive to large-scale driving. Unless the energy injection rates on small scales are much higher than the energy injection rate on large scales, the large-scale driving will always dominate. Therefore, even in the presence of significant stellar feedback in the Galactic plane,  large-scale turbulent driving might be the dominant mechanism. The finding of a relatively uniform SPS slope for the Milky Way resembles the situation with the \HI\ in the Small Magellanic Cloud where  \citet{szotkowski19} found a very uniform SPS slope across the entire galaxy. 

\citet{hennebelle12} showed that the kinetic energy transfer rate measured for the population of molecular clouds traced by $^{12}$CO showed no variation with spatial scale ranging from $\sim0.01$ to $\sim100$ pc. 
They concluded that this result suggests that molecular clouds are part of the same turbulent cascade as \HI. From this point of view, the relatively uniform \HI\ SPS slope may not be so surprising -- if molecular clouds of varying levels of star formation and stellar feedback sources do not show significant changes of their internal turbulent properties, the \HI\ may be also not be strongly affected by small-scale turbulent driving.

In summary, while the interpretation of the \HI\ SPS remains complicated, some trends are starting to emerge by analyzing large surveys with consistent methodologies. In moving forward, it is essential to develop a consistent and uniform approach for handling data, as well as more sophisticated methods for disentangling geometry, phase, and turbulent mode effects.

\subsection{3-D Distribution }
\label{subsec:3d_distribution}

\subsubsection{Radial Distribution}
By nature of the large-scales involved, our overall understanding of the 3-D distribution of the Milky Way in \HI\ has not changed much since the excellent review by \citet{kalberla09}.   They carefully explored the radial distribution, the warp and the scale height of the warm disk.  Key characteristics included an asymmetric warp beyond $R\approx 9$ kpc and a rapidly flaring disk beyond about $R>18$ kpc.


Like most \HI\ rich galaxies, the Milky Way has an extended \HI\ disk.  What is the true extent of the \HI\ disk and what form does it take?  It is difficult to associate a distance for the outer Milky Way because observational methods are, in general, limited to kinematic distances.  With the assumption of a mostly flat rotation curve \cite{kalberla08} found that the \HI\ disk extends beyond $R\sim 30$ kpc, but the \HI\ surface density there is low: $N_H\sim 1.5 \times 10^{19}~{\rm cm^{-2}} $ or $0.1~{\rm M_{\odot}~pc^{-2}}$.  At even larger galactocentric radius, the disk likely extends as far as 60 kpc but with a very high velocity dispersion ($\sigma_v =74~{\rm km~s^{-1}}$), which cannot be thermal and therefore must suggest a clumpy distribution with a high cloud-to-cloud velocity dispersion.   A large radial extent to the \HI\ disk has also been observed in other galaxies, for example NGC 2997 \citep{pisano14} shows clumps of \HI\ up to 60 kpc.  

The tail end of the \HI\ radial distribution, however, is at odds with previous theoretical estimates of the column density cutoff due to ionization from the CMB \citep[e.g.][]{dove94,maloney93}, which suggest that the \HI\ column density should drop precipitously after $R\sim 35~{\rm kpc}$.   \citet{bland-hawthorn17} revisited the extended \HI\ disk models including the effects of non-uniform gas distribution or reducing the scale height.  They showed that if the \HI\ distribution was compressed into a narrower scale-height,  increasing the mean volume density by a factor of 100 to $n_H \sim 10~{\rm cm^{-3}}$, the disk could survive as \HI\ with $N_H > 10^{18}~{\rm cm^{-2}}$ to $R>55$ kpc.  But the radial extent of the disk is strongly dependent on the assumed mean volume density; reducing to $n_H  \sim 1~{\rm cm^{-3}}$ decreases the disk radius at the same sensitivity limit to $R\sim 45$ kpc. 

Explaining the Milky Way's apparently large, low column, radial extent relies on both increasing the mean volume density and also explaining the high velocity dispersion.  The latter can be possibly attributed to the discrete extended \HI\ line wings like those observed by \citet{kang07} and tentatively associated with old supernova remnants. The evidence for the large volume density predicted by \citet{bland-hawthorn17} is not as clear.  Certainly, we know from \citet{strasser07} and \citet{dickey09} that dense material exists at radii $15<  R<25$ kpc with estimated number densities of $n_H \approx 0.7 - 2.4 ~{\rm cm^{-3}}$.   

It is worth noting that there is some direct evidence  for the \HI\ extending at least as far as $R=23$ kpc.  This comes from the  interaction between the Magellanic Leading Arm and the Milky Way disk \citep{mcgriff04,price-whelan19} at $R=23$ kpc, which has resulted in an orphan star cluster associated with the Magellanic Leading Arm and formed from the interaction with the disk.   \citet{nidever19} used the full three-dimensional velocity vector and star cluster's orbit to predict the gas density of the disk through which the cluster passed.  This interesting example provides an unusual method to estimate the  density of the \HI\ disk at the  midplane at $R=20$ kpc: $n_H = 6.0 \times 10^{-2}~{\rm cm^{-3}}$. 

The CNM and WNM have different spatial extents.  Unlike the warm gas, the scale height and extent of the CNM disk has not been well characterized.  For the most part this has been an observational limitation; \HI\ emission is dominated by warm gas and the only places where the profiles are simple enough to allow CNM-WNM decomposition is at high latitudes and dominated by the local gas.  Nonetheless, the sheer number of \HI\ absorption measurements enabled by the Galactic plane surveys \citep[CGPS, VGPS, CGPS, THOR;][]{taylor03,stil06a,mcgriff05,wang20a}, has given some indications of the extent of the cold disk.

The cold \HI\ disk extends to at least 25 kpc and shows a distinct pattern in longitude-velocity space that nominally traces the spiral arm distribution \citep{strasser07}.  
As we already discussed in Section 5.2,
\citet{dickey09} showed that the 
CNM fraction remained mostly flat with Galactic radius out to at least 20 kpc, with the assumption that the average temperature of the cold component does not change. Lately,  \cite{dickey22}  has revisited that result  by stacking 175  \HI\ absorption spectra towards extragalactic sources from the new Galactic ASKAP (GASKAP) Pilot Survey and found that the 2009 result stands with a constant CNM fraction
out to $R\sim 40$ kpc. These results are somewhat surprising; the outer Galaxy environment has a much  metallicty (possibly $<1/10$ the Solar circle gas phase O/H abundance \citep{wenger19}) and the balance of heating and cooling is likely quite different.

\subsubsection{Scale Height}
The scale height of the \HI\ emission layer and its variations with Galactocentric radius exterior to the solar circle have been  well measured by \citet{kalberla08} and summarized in the \citet{kalberla09} review.   \citet{kalberla09}  described the radial dependence of the full-width at half-max, FWHM, of the emission layer between $R=5~{\rm kpc}$ and $R=35~{\rm kpc}$ as $f= f_0 e^{(R-R_0)/R_0}$, where $f_0 = 300~{\rm pc}$ and $R_0 = 9.8$ kpc.   Interior to the solar circle, from $4 \lesssim R  \lesssim 8$ kpc, the density, $n(z)$ of the  \HI\ disk in the inner Galaxy has long been characterized by a dual Gaussian with FWHM of 212 pc and 530 pc, with an additional low density exponential tail \citep{lockman86,dickey90,savage09}.  The full distribution has a FWHM of 430 pc but \citet{lockman84} also noticed that inside about $R=2 $ kpc, the scale height appeared to decrease.  This was confirmed in detail by \cite{lockman16}, who showed that from $R \approx 2.1$ kpc on the negative longitude side of the Galactic center to $R \approx 2.4$ kpc on the positive longitude side the large scale height layer is missing and the disk is fit very well by a single Gaussian with FWHM $\sim 125$ pc.  This diminished disk is matched with a dramatic reduction in the \HI\ surface density in the inner 2.5 kpc of the Galaxy.  At $R = 3.5 - 3.75$ kpc the surface density is $\Sigma_{HI} = 3.4~{\rm M_{\odot}~pc^{-2}}$, dropping to $\Sigma_{HI} = 0.4~{\rm M_{\odot}~pc^{-2}}$ in the annulus from $1.5$ kpc to $1.75$ kpc.  The cause of the  significant drop in \HI\ surface density is not clear, with possible causes related to the conversion of \HI\ to molecular material, evacuation due to Galactic centre activity or simply an observational effect of non-circular orbits \citep{lockman16}.

The scale height of the \HI\ emission disk may also vary in the outer Galaxy.  Working from a flat rotation curve \citet{levine06} transformed the brightness temperature distribution of \HI\ emission, $T(l,b,v)$ to an estimated number density distribution $n(R, z, \phi)$ and found variations in the scale height of 15-20\% correlating with variations in the \HI\ gas surface density $\Sigma_{HI}$ of 30-50\%.  Using a similar technique, but fitting for the rotation curve as well, \citet{kalberla07} showed the same general trends. The conclusion from these works is that the gas disk is thinner where the surface density is higher, and that these variations correlate with spiral structure.  It should be noted, however, that these conclusions are based on nearly flat, azimuthally smooth rotation curves.  We know that dense spiral arms can drive up to 10\% deviations from a flat rotation curve \citep{burton71,mcgriff16,peek22}, which can drive systematic variations in distance dependent properties.  The conclusions of varying \HI\ scale height correlating spiral arms should therefore be viewed with some caution.

While the distribution of the \HI\ emission layer has been well characterized across the Galaxy, the distribution of the cold \HI\ disk has been much more difficult to constrain.  Focusing on the solar neighborhood, \citet{crovisier78} pioneered a technique to ascertain the scale height of the cold \HI\ disk at the solar circle from observations of \HI\ in absorption. By comparing the radial velocities in a catalog of absorbing \HI\ clouds at $|b|>10^\circ$ (isolating only the local clouds) to those expected for clouds in a differentially rotating disk, \citet{crovisier78}, and later \citet{belfort84}, estimated the mean distance of local \HI\ clouds from the plane, $\langle|z|\rangle\approx100~\rm{pc}$.  Both \citet{crovisier78} and \citet{belfort84}  found that $\langle|z|\rangle$ increased at higher latitudes ($|b| \gtrsim 20^\circ$), interpreted as a result of the low abundance of \HI\ clouds very near to the Sun.  For a Gaussian distribution $\langle|z|\rangle=100~{\rm pc}$ corresponds to a FWHM of almost 300 pc. 

\begin{figure}[h]
	\centering
	\includegraphics[width=\textwidth]{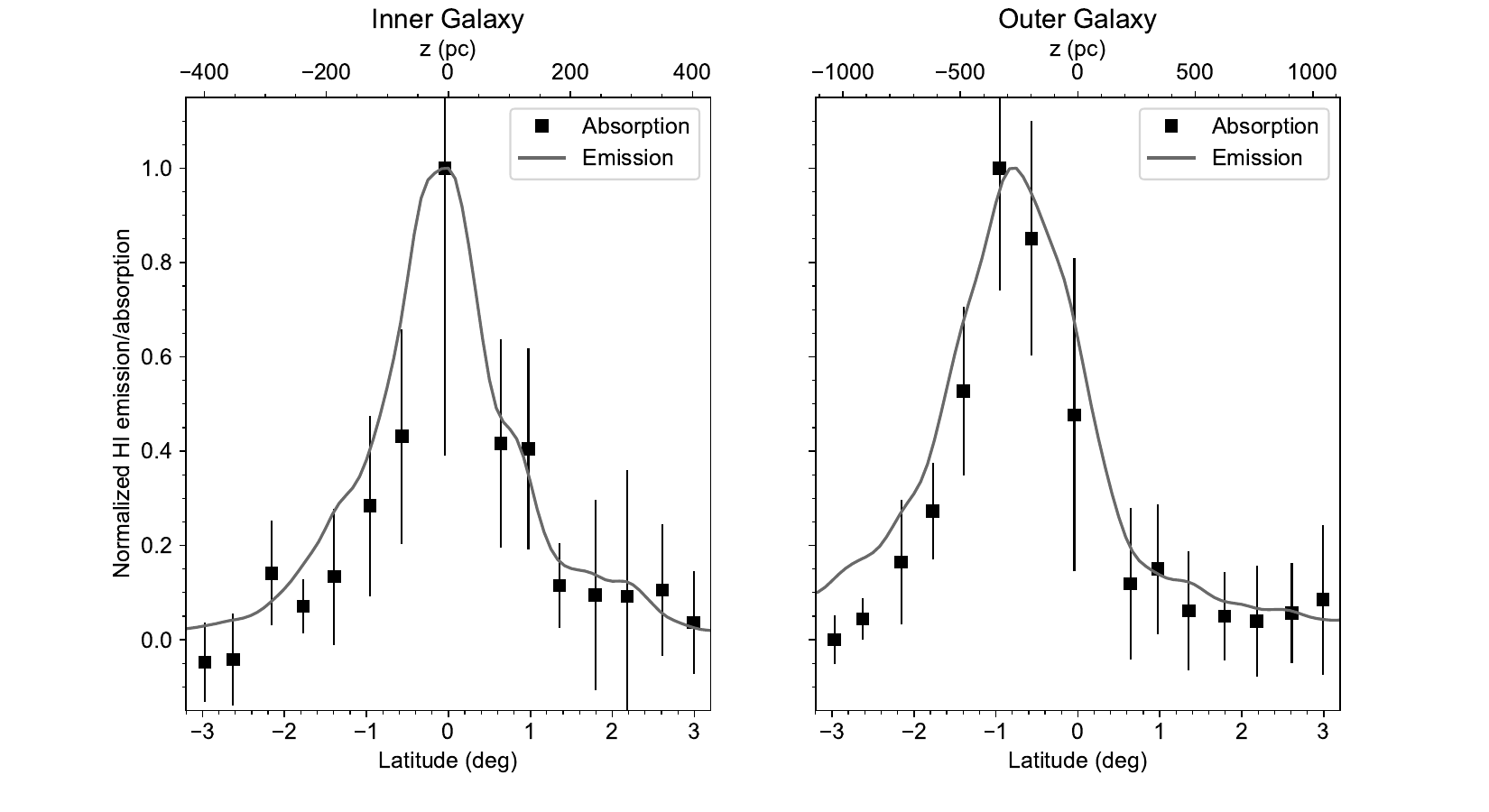}
	\caption{Normalized distributions of average fractional absorption, $\left<1-e^{-\tau}\right>$, and average emission, $\left<T_b\right>$, versus Galactic latitude between $l=339^{\circ}$ and $l=342\arcdeg$ for velocities close to the inner Galaxy tangent point at $R\sim 2.8$ kpc (left) and in the outer Galaxy at $R\sim 12.5$ kpc (right).  The upper axis in $z$ is for an assumed distance of $d=7.3$ kpc and $d=20$ kpc for the tangent point and assuming a flat rotation curve with $R_{\odot}=8.31$ kpc and $\Theta_0 = 240~{\rm km~s^{-1}}$.   Absorption data were obtained from \citet{dickey22} and emission data from GASS \citep{mcgriff09,kalberla10}.  }
	\label{fig:HI_cold_z}
\end{figure}

The irregular aerial sampling of \HI\ absorption through the low Galactic latitudes, which are most useful for probing the scale height of the cold disk, has been a limitation for understanding the CNM disk structure away from the solar neighborhood.   \citet{dickey09} used the Galactic Plane surveys to produce a projection of \HI\ opacity as a function $R$ and $\phi$ outside the Solar circle.  Recently,  the \citet{dickey22} measurements of \HI\ absorption towards 175 extragalactic continuum sources in the GASKAP Pilot region have provided the first indications of the scale height of the CNM disk in several regions away from the solar circle.  \citet{dickey22} stacked \HI\ absorption spectra from the region $338^{\circ} \leq l \leq 342^{\circ}$, focusing on regions exterior to the solar circle and in the inner galaxy at the sub-central point where the line-of-sight is tangent to a circular orbit around the Galactic center and the \HI\ reaches its terminal velocity \citep[e.g.][]{mcgriff16}.  These measurements, together with Solar neighborhood measurements \citep{crovisier81,belfort84}, are starting to give a clear picture of variations in the height cold \HI\ disk with Galactocentric radius.

\begin{figure}[h]
	\centering
	\includegraphics[width=3in]{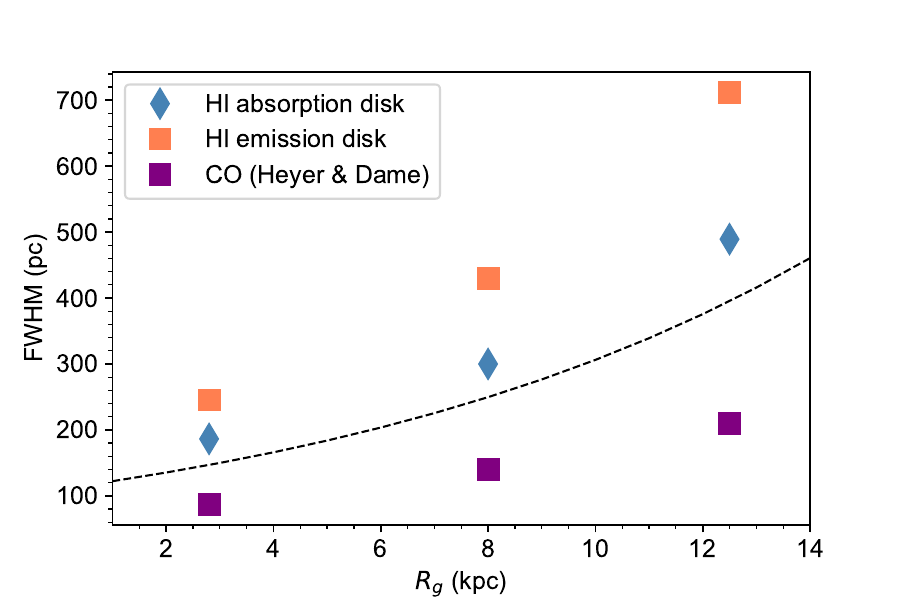}
	\caption{Scale height comparison of cold and warm \HI\ with CO disk.  Blue points are the FWHM of the absorption disk, orange points are FWHM of the emission disk and purple points are the FWHM of the CO from \citet{heyer15}  and the dashed line is the \cite{kalberla09} fit for $n(z)$ based on the assumption of hydrostatic equilibrium.  Absorption data are from \citet{crovisier81} and \citet{dickey22}, emission FWHM were calculated from Gaussian fits to HI4PI data \citep{hi4pi-collaboration16}. }
	\label{fig:fwhm_vs_r}
\end{figure}

The $z$ structure of the emission and absorption disks in the inner and outer Galaxy are compared in Figure~\ref{fig:HI_cold_z}. In this Figure we have compiled the \citet{dickey22} data, showing the comparison of the averaged \HI\ absorption $\left<1-\exp(-\tau)\right> \Delta v$ to the averaged \HI\ emission $\left<T_b\right>$ as a function of latitude in a region in the inner  ($R\sim 2.7~{\rm kpc}$) and outer ($R\sim 12.8~{\rm kpc}$) Galaxy.  The inner Galaxy region shows only the absorption and emission  at the terminal velocity $-145~{\rm km~s^{-1}} < v_{LSR} < -125~{\rm km~s^{-1}}$, where the distance is geometrically defined at $d\sim 7.8~{\rm kpc}$ for  $R_{\odot} = 8.3~{\rm kpc}$.  The outer Galaxy region shows absorption and emission between  $15~{\rm km~s^{-1}} < v_{LSR} < 35~{\rm km~s^{-1}}$,  where for a flat rotation curve $d\approx 20$ kpc and $R\sim 12.5$ kpc.    Fitting to the normalized $z$ distributions, we find that the FWHM of the absorption disk is 185 pc at $R\sim 2.8~{\rm kpc}$ and 490 pc at $R\sim 12.5~{\rm kpc}$, while the inner and outer Galaxy emission FWHM's are 245 pc and 710 pc, respectively.  The fits are consistent with \citet{dickey22}, yet we also confirm the classic view that for the inner Galaxy the emission is better fit with two Gaussians of FWHM 110 and 340 pc,  while the low resolution of HI4PI limits the outer Galaxy fit to a single Gaussian component.  The key conclusions from the comparison are: the absorption disk is narrower than the emission disk at all Galactic radii and the CNM and emission disks are narrower in the inner Galaxy than in the outer Galaxy.  When combined with the \citet{crovisier81} a picture develops of the cold \HI\ disk increasing in thickness from a FWHM of $\sim 185$ pc at $R\sim 2.8~{\rm kpc}$, through 300 pc around the Sun, to $\sim 490$ pc at $R\sim 12.5$ kpc (see Figure~\ref{fig:fwhm_vs_r}).  These conclusions are  not necessarily surprising given the higher pressure for the inner Galaxy, but suggest that the cold \HI\ gas disk more closely matches the $H_2$ disk than the warm \HI.

Ideally, analyses like Dickey et al (2009, 2022)\nocite{dickey09,dickey22}  will be extended across the Galactic disk to help us realize an azimuthally resolved view of the disk scale height.  If non-circular motions could be accounted for with future stellar derived mass models for the Milky Way we might also hope to be able to incorporate those and overcome problems related to the rotation curve.

\section{FUTURE SURVEYS OF ATOMIC HYDROGEN IN THE MILKY WAY}
The field of atomic hydrogen in the Milky Way is once again poised for an observational revolution.  The new SKA Pathfinders, ASKAP and MeerKAT, are providing highly sensitive interferometric data over wide-fields, helping to bridge the gap between the limited area, low sensititivity interferometric surveys with $\sim$ arcminute resolution and the large area, high sensitivity single dish surveys with $3-16$ arcminute resolution.  MeerKAT has a one degree field-of-view with outstanding surface brightness sensitivity at sub-arcminute angular resolution. With its  surface brightness sensitivity, MeerKAT 
will make excellent contributions to highly sensitive imaging over areas of 10's of degrees.  We can envisage that MeerKAT will lead the way with deep, 10\arcsec\  imaging of high-latitude \HI, with applications towards phase decomposition of \HI\ emission and studies of turbulence.

ASKAP makes use of phased-array technology to extend its field-of-view to $\sim 25~{\rm deg^{-2}}$ with a surface brightness sensitivity at 30\arcsec\ that is similar to previous interferometric \HI\ surveys but 300 times faster \citep{hotan21}.   The GASKAP-HI survey is  specifically targeting \HI\ in the Milky Way at high angular ($\sim 30\arcsec$) and spectral resolution ($ \sim 0.4~{\rm km~s^{-1}}$) with 
an order of magnitude improvement in sensitivity and angular resolution on the Galactic Plane, Galactic Center and a large swathe of high latitude sky covering the Magellanic Clouds and Stream plus Milky Way foreground.   GASKAP-HI  will jointly deliver \HI\ emission images and catalogs of \HI\ absorption.  These data will be outstanding for joint studies of the CNM and WNM distribution with galactocentric radius and in $z$ height.  

\HI\ absorption surveys at high Galactic latitudes  have been slow to carry out and almost exclusively targeted. For example, the MACH survey measured \HI\ absorption in the direction of just 42 background sources in 50 hrs of observation to trace the latitude dependence of absorption properties.  The GASKAP-HI high latitude survey GASKAP-HI will obtain $\sim 150$ absorption spectra of similar quality to MACH in every $25~{\rm deg^2}$ field after 30 hours of integration.    In the Galactic foreground of the Magellanic Stream, GASKAP-HI should yield over 7000 new \HI\ absorption measurements over a range of Galactic latitudes. The spatial sampling of this ``\HI\ absorption grid" will help trace variations in spin temperature and column density over scales of a parsec or less.   

The next major advance in \HI\ studies of the Milky Way will come when the spatial sampling of an all-sky irregular grid of \HI\ absorption finally reaches the effective angular resolution that we currently have with HI4PI in emission. The full SKA, when operational in $\sim$ 2028, will extend Milky Way \HI\ studies further in both emission and absorption.  Together the pathfinders and ultimately the SKA, will  be able to provide a catalog several orders of magnitude larger than \bighicat, with values of optical depth, spin temperature and column density  determined from a dense grid of absorption measurements, finally allowing us to realize the goal of having all (Southern)-sky cubes of the CNM with an effective resolution of $\lesssim 0.25$ deg.    

Two other telescopes that will enormously improve the sensitivity of \HI\ surveys are: FAST for \HI\ emission  and ngVLA for \HI\ absorption.  By virtue of its 500 m collecting area, FAST can reach brightness temperature sensitivity limits better than any other telescope in the world at a few arcminutes angular resolution.  The ngVLA, which if designed to specification, will be superb for measuring \HI\ absorption to deduce the properties of the WNM.  ngVLA should produce absorption spectra with about 60 times lower noise for a given observation time than 21-SPONGE, which is the most sensitive comprehensive survey of \HI\ absorption.   If the spectral baselines are stable the ngVLA should produce observations at $\sigma_{\tau} <10^{-5}$ towards bright ($\sim 1 - 10$ Jy) sources with minimal telescope time. 

As we move into the realm of  ``big data" for \HI\ it is no longer possible to look at each cube and fit each spectrum by hand.  It will be essential to bring new techniques to bear for extracting physical properties from the data as well as characterizing the morphological interpretation of multi-dimensional data.  Some of these new techniques have recently been tested and applied to new surveys \citep[e.g. Gausspy, ROHSA, AstroHOG, RHT, CNN][]{lindner15,marchal19,soler19,clark15,murray21}.  The requirements for new tools and techniques to analyze the spectrally and spatially complex  \HI\ datasets will be essential to  further our understanding of \HI\ in the Milky Way.  Applying these techniques to synthetic observations of simulations will help us to infer astrophysical quantities from  the observational \HI\ data.

\subsection{Future complementary data}
New \HI\ data will be complemented by other datasets capable of tracing the dust, molecular gas and magnetic field structure of the ISM with better resolution in all dimensions.  We already starting to see  the  power of linking \HI\ datasets with the 3-D dust distribution  within a few hundred parsecs of the Sun   \citep[e.g.][]{lallement19,green19,leike20,vergely22}. For example, \citet{panopoulou20} used the association with dust to show that there are 2 to 3  kinematically distinct \HI\ components along most high Galactic latitudes lines-of-sight. \citet{tchernyshyov17} and \citet{tchernyshyov18} used the 3-D dust together with \HI\ for 4-D (3 spatial plus 1 velocity) information about the gas distribution.  While these dust maps are currently limited to local gas, future infrared surveys with the Nancy Grace Roman Space Telescope could be used to extend the 3-D dust distribution within the Galactic plane to much further distances.  Associating \HI\ with dust to derive distances and trace 4-D \HI\ information will open new opportunities to use \HI\ for spiral structure, physical gas properties and many other areas.

Essential to interpretation of future  observational data will be future simulations at sub-parsec resolution over the full disk of a galaxy to trace the micro-physics visible through our new observational datasets.  
Simulations have a particular advantage as they can turn on and off individual heating and cooling processes and introduce different feedback channels.  By comparing the simulated mass fractions across different phases with observations we can gain insight into the importance of various feedback channels.  From this point of view, a database of simulations made consistently while varying different input parameters would be very useful for future investigations.
With thousands of \HI\ absorption spectra to be provided by future surveys, we envision that both observations and simulations can start to focus on individual regions (e.g.\ Galactic center vs. outer regions, the \HI\ warp, off-plane, etc.), instead of the entire Milky Way, to investigate the mass flow of \HI\ across different phases and how this changes with local conditions.

\begin{summary}[SUMMARY POINTS]
\begin{enumerate}

\item The CNM is ubiquitous in the Milky Way. The CNM temperature distribution 
peaks at 50-200 K and is broad, likely due to the sampling of both dense and diffuse environments.
Some evidence exists that only colder and denser CNM goes on to form \htwo.
The CNM temperature decreases with $A_V$: higher-$A_V$ regions have a lower $T_s$,
while lower-$A_V$ regions have
a higher $T_s$.

\item The average CNM mass fraction in the Milky Way is $\sim40$\%. 
The CNM fraction increases with Av due to the increase of the mean density. Local density plays a strong role in constraining the CNM fraction and spin temperature. The morphology of the CNM, as observed in emission, is highly filamentary and is correlated with the orientation of the magnetic fields.

\item The UNM remains difficult to constrain observationally because it requires highly sensitive absorption measurements.
The most robust estimate to date for the average UNM mass fraction is $\sim 20$\%, in agreement with predictions from        
recent numerical simulations.
The morphology, spatial distribution, and other properties of the UNM still remain as open questions.

\item The WNM peak of the \HI\ temperature distribution remains observationally unconstrained from \HI\ absorption
observations due to its very low optical depth. The likely reason is that the WNM has
a higher excitation temperature than what has been assumed over the last few decades.
The spatial distribution of the WNM in relation to the CNM and UNM needs better constraints.
The average WNM mass fraction is $\sim40$\%, even the densest ISM regions have some WNM.
Shells and superbubbles are important for setting the \HI\ emission morphology and create
\HI\ structures as broken remnants of shells and supershells driven away from
the disk.

\item The interpretation of the spatial power spectrum of \HI\ emission depends on many effects.
Large areas of the Milky Way appear to have a relatively uniform SPS slope.

\item The cold disk of the Milky Way extends to at least three times the solar radius.  Throughout the disk,  the CNM  has a scale height that is intermediate between that of the WNM and the molecular gas.
\end{enumerate}
\end{summary}

\begin{issues}[FUTURE ISSUES]
		
	Throughout the compilation of this review we encountered countless questions for future work.  It is impossible to describe them all here, but we have chosen to list below some of the biggest questions that we might reasonably hope to answer within the next decade through savvy combination of multi-wavelength observational and simulated datasets 
	and new techniques.   
	
\begin{enumerate}

	\item New surveys of \HI\ absorption will expand the number of \HI\ absorption detections by a factor more than 100 to help us to answer questions about  how  the spin temperature changes with location around the Galaxy, hopefully  revealing what fraction of the observed \HI\ is in the UNM and how  this varies with environment.  These extensive databases of \HI\ absorption should 
	show  how the \HI\ excitation processes 
	and the heating and cooling rates 
	change with environment, e.g.\ within the star-forming disk versus the halo, and inner versus outer disk regions.
	
	\item Projects combining stellar and dust distances to produce a complex model of the 3-D structure of the Milky Way's ISM are  currently underway.  Coupling a 3-D ISM model with \HI\ spectra  will help to break some of the current distance velocity degeneracies that trouble \HI. With this we might hope to be able to answer: How do the \HI\ properties correlate with large-scale Galactic structure?  Are there opacity or scale height variations with $R$ and azimuth?   Can \HI\ be routinely linked to dust to provide 3-D spatial information together with the velocity information? The 3-D dust and stellar distribution models will also help researchers to create a 3-D UV radiation field to  understand how the \HI\ properties change with local UV radiation field or column density.
	
	\item By combining new high spatial and spectral resolution \HI\ surveys  with machine learning and other spectral decomposition techniques, future work may be able to investigate the CNM and WNM separately everywhere in the Galaxy.  Combining fully decomposed atlases of the CNM and WNM with ever improving statistical descriptors of morphology (e.g. Rayleigh statistics \citet{soler20}) we hope that researchers will be able to understand the differences in structure between the phases and whether magnetic fields are a dominant contributor to the structure of the \HI\ in both or in just the CNM. 
	
	\item With the revolution of large-area, high-resolution \HI\ observations, \HI\ will remain  a unique tracer for observationally determining  the relative
	importance of various turbulent driving modes in diverse ISM environments. 
	As numerical simulations become increasingly complex and  move beyond prescribed turbulence driving to directly include 	various channels of stellar feedback they will require the detailed understanding of the 
	effects these modes have on the surrounding ISM that will be provided by future statistical studies of \HI.
	
	\end{enumerate}
	
Perhaps the most important question is: {\bf What lessons can we take from the Milky Way to help us understand other galaxies? } Over the next five years extragalactic \HI\ surveys with the Jansky VLA, ASKAP and MeerKAT will start to reveal nearby galaxies with the spatial resolution and number of \HI\ absorption spectra that we have for the Milky Way in this review.  Through comparison of Milky Way \HI\ with the properties of \HI\ measured in nearby galaxies we will start to probe how the \HI\ phases responds to metallicity and vastly different UV radiation fields.   
We started this review noting that it is only in the Milky Way that we can resolve the sub-parsec scales needed to understand the transition from the WNM to the CNM and its role as a throttle in  star formation.  It is not obvious that we yet understand the full details of that transition, but it it is clear that the physics is rich and by extending detailed \HI\ studies from the Milky Way to other galactic environments we might just reach an answer.

\end{issues}

\section*{DISCLOSURE STATEMENT}
The authors are not aware of any affiliations, memberships, funding, or financial holdings that
might be perceived as affecting the objectivity of this review. 

\section*{ACKNOWLEDGMENTS}
We have benefited enormously from conversations with S.\ Bialy, J.\ M.\ Dickey, M.\ Gong, P.\ Hennebelle, C.-G. Kim, A.\ S.\ Hill, A.\ Marchal, M.-A.\ Miville-Desch\^{e}nes, C.\ Murray, \& H.\ Nguyen.  We are very grateful to Eve Ostriker for many suggestions that have improved the early version of the manuscript. We are grateful to Amit Mittal for providing Figure 8. The authors acknowledge the Interstellar Institute's program ``With Two Eyes" and the Paris-Saclay University's Institut Pascal for hosting discussions that nourished the development of the ideas behind this work. The authors are grateful to D.\ McConnell for  significant comments on drafts of this review. NMMc-G acknowledges funding support through an Australian Research Council Laureate Fellowship (project number FL210100039) funded by the Australian Government.  SS and DRR acknowledge support provided by the University of Wisconsin - Madison
Office of the Vice Chancellor for Research and Graduate Education with funding from the Wisconsin Alumni Research Foundation. This work was supported by the NSF Award AST-2108370 and the National Aeronautics and Space Administration under Grant No. 4200766703.

%
%
%
%
%

\bibliographystyle{ar-style2}

\bibliography{references}


\end{document}